# Tunable and Transferable Diamond Membranes for Integrated Quantum Technologies

Xinghan Guo,[†] Nazar Delegan,[†,‡] Jonathan C. Karsch,[†] Zixi Li,[†] Tianle Liu,[¶] Robert Shreiner,[¶] Amy Butcher,[†] David D. Awschalom,[†,‡,¶] F. Joseph Heremans,[†,‡] and Alexander A. High[*,†,‡]

[†]*Pritzker School of Molecular Engineering, University of Chicago, Chicago, USA*

[‡]*Center for Molecular Engineering and Materials Science Division, Argonne National Laboratory, Lemont, USA*

[¶]*Department of Physics, University of Chicago, Chicago, USA*

*E-mail: ahigh@uchicago.edu

**Abstract**

Color centers in diamond are widely explored as qubits in quantum technologies. However, challenges remain in the effective and efficient integration of these diamond-hosted qubits in device heterostructures. Here, nanoscale-thick uniform diamond membranes are synthesized via "smart-cut" and isotopically ($^{12}$C) purified overgrowth. These membranes have tunable thicknesses (demonstrated 50 nm to 250 nm), are deterministically transferable, have bilaterally atomically flat surfaces ($Rq \leq 0.3$ nm), and bulk-diamond-like crystallinity. Color centers are synthesized via both implantation and in-situ overgrowth incorporation. Within 110 nm thick membranes, individual germanium-vacancy (GeV$^-$) centers exhibit stable photoluminescence at 5.4 K and average optical transition linewidths as low as 125 MHz. The room temperature spin coherence of individual nitrogen-vacancy (NV$^-$) centers shows Ramsey spin dephasing times ($T_2^*$) and Hahn echo times ($T_2$) as long as 150 µs and 400 µs, respectively. This platform enables the straightforward integration of diamond membranes that host coherent color centers into quantum technologies.

# Introduction

Color centers in diamond are a leading platform in quantum networking and sensing due to their exceptional coherence times,[1,2] robust spin-photon interfaces,[3,4] and controllable interactions with local nuclear- and reporter-spin registers.[5–7] Color centers have been utilized in many landmark experimental demonstrations, including deterministic entanglement,[8] multi-node quantum networking,[9] nanoscale NMR spectroscopy,[10] and memory-enhanced quantum key distribution.[11] These advances can be accelerated into scalable technologies with the ability to create high quality, nanoscale-thick diamond membranes as elements in hybrid devices. Ideally, in the next generation of quantum technology, diamond will simply be another functional layer in heterostructures that can include non-linear/magnetic/acoustic materials, on-chip detectors, superconductors, nanophotonics, and microfluidics. Such improved integration can increase entanglement efficiency,[9] master control of phonons and photons,[11–13] enable on-chip frequency conversion,[14] improve interfaces with other quantum systems and sensing targets,[15,16] and create new opportunities to engineer quantum states. However, the material properties of diamond create fundamental difficulties for heterostructure integration. Specifically, high quality, single crystal heteroepitaxial growth of diamond thin films remains challenging despite recent progress.[17,18] In response, a variety of processing and integration schemes have been developed to derive low dimensional structures out of bulk diamond.[19–24] While promising, a scalable, high yield approach that enables full heterostructurelike integration of diamond while maintaining bulk-like properties – specifically, crystallinity, surface roughness, and color center coherence – is still lacking. Here, we report the efficient synthesis and manipulation of ultra-thin diamond membranes suitable for next-generation applications in quantum information science (QIS). Our membrane fabrication procedure is based on a "smart-cut" technique,[23,25–28] in which He$^+$ implantation creates a sub-surface graphitized layer that can be electrochemically etched, in conjuncture with a plasma enhanced chemical vapor deposition (PE-CVD) overgrowth optimized for the synthesis of high quality, in-situ doped, and isotopically purified diamond. The process allows detachment of arbitrarily large diamond membranes with smooth interfaces without relying on reactive-ionetching (RIE) undercut processes. While previous works have demonstrated the effectiveness of "smart-cut" and overgrowth for diamond membrane creation, our process realizes three critical advancements to the state-of-the-art. Firstly, we demonstrate isotopically controlled and nitrogen δ-doped diamond membrane structures with atomically flat surfaces and high crystal quality showing unprecedentedly narrow Raman linewidths.[28,29] Secondly, we demonstrate a novel dry-transfer technique which enables clean and deterministic membrane placement on arbitrary substrates. Membranes transferred in this way are free from premature detachment and unwieldy curvatures caused by built-in strain between the graphitized and overgrown layers.[30] As such, these membranes can be integrated with other material platforms and functional structures. Thirdly, we report that these structures, even at thickness ≤150nm, are suitable hosts for germanium vacancy (GeV$^-$) and nitrogen vacancy (NV$^-$) centers created via implantation (in both cases) and overgrowth (NV$^-$ only). Specifically, we show that the GeV$^-$ centers exhibit stable and coherent emission at 5.4 K, while the NV$^-$ centers show bulk-like spin coherence properties at room-temperature. Additionally, we demonstrate that doping of other group IV color centers such as silicon-vacancy (SiV$^-$) and tin-vacancy (SnV$^-$) is also viable.

## Results and discussion

### Diamond Membrane Synthesis

The step-by-step fabrication procedure of the diamond membrane quantum platform is demonstrated in Figure 1a-e. The process starts with a low energy (150 keV) He$^+$ implantation ($5 \times 10^{16}$ cm$^{-2}$) into diamond substrates, as shown in Figure 1a. This step forms a depth-localized graphitized underlayer ≈410 nm deep[31] via damage-induced phase transition of the carbon bonds from sp$^3$ to sp$^2$. Unlike RIE-based undercut approaches,[21,22] the top diamond layer maintains a uniform thickness and flatness throughout processing (see Figure S1b and Figure S3a in[32]). The substrates are then subjected to a multi-step anneal with a maximum temperature of 1200 °C (see section 1.1 in[32]). The high temperature allows for the mobilization and subsequent annihilation of implantation-induced crystal damage in the top layer[33,34] as characterized via Raman spectroscopy (see section 2.2 in[32]). However, this process is imperfect, resulting in the top layer remaining unsuitable as a host for highly coherent color centers.

To fully mitigate the impact of implantation damage and achieve pristine crystal quality in the membranes, we follow the "smart-cut" with homoepitaxial PE-CVD of an isotopically engineered diamond thin film overlayer as shown in Figure 1b. During growth, the hydrogen to methane flow rate ratio is kept at 0.05% to ensure a morphology preserving step-flow growth regime.[1,35] The growth rates herein were $6.2(4)$ nmh$^{-1}$ to $9.3(8)$ nmh$^{-1}$ for 700°C to 500°C heating plate temperature, respectively (see section 1.2 in[32]). The process can be performed with higher growth rates to efficiently achieve thicker structures. In this work, we maintained low rates compared to other works[23,26] to ensure a more accurate depth-localization of dopant layers, i.e., $\delta$-doping precision.[1]

We employed two distinct strategies for point defect creation within the overgrown layer. A subset of the membranes had a ≈2 nm $\delta$-doped layer of $^{15}$N grown in, as schematized on the right of Figure 1c. These membranes underwent electron irradiation and subsequent annealing to form a $\delta$-doped NV$^-$ layer (see section 1.2 in[32]). Other overgrown diamonds were ion-implanted with Ge$^+$, Si$^+$, Sn$^+$, and N$^+$ as seen on the left of Figure 1c. These membranes were subjected to another identical multi-step anneal to form optically-active point defects via vacancy mobilization and annihilate implantation-induced crystal damage as much as possible.[33,36]

To realize a fully integrable diamond platform, we have engineered a high yield, controllable process to lithographically pattern membranes with desired shapes and sizes into the overgrown film, and subsequently transfer them individually onto other substrates/devices. The left of Figure 1f shows inductively coupled plasma (ICP)-defined square-shaped membrane arrays (200 µm side length) used in this work. Each step of the membrane definition and transfer utilizes established techniques and can be done in a cleanroom with standard equipment (see section 1.3 and 1.4 in[32]). The size and shape of the membranes are fully defined and can be tailored to specific applications, with the maximum size only limited by the substrate dimensions. We fabricate membranes that are 200 µm by 200 µm as a demonstration of sufficient size for photonics integration.[11,13,21] Manipulation of the membranes starts with an electrochemical (EC) etching of the graphitic underlayer as shown in Figure 1d. The substrate is placed in deionized (DI) water, with two electrodes in the vicinity (≤25 µm) of the target membrane. These electrodes generate the EC potential necessary to selectively etch the graphitic layer(see section 1.3 in[32]). Critically, in contrast to previous studies,[26,30] a small portion of the underlayer is left unetched, creating a tether (selectively unetched diamond underlayer) that prevents premature membrane detachment before the dry transfer. The overall efficiency to transfer membranes onto a carrier wafer is greater than 80%, mainly limited by human error during the EC etching. Therefore, from a single 3 mm by 3 mm diamond substrate, we can derive more than 45 (200 µm × 200 µm) functional membranes.

Our transfer process, depicted in Figure 1e, draws inspiration from those utilized in van der Waals heterostructure fabrication.[37] This process starts with a polydimethylsiloxane (PDMS) polycarbonate (PC) stack mounted on a micropositioner with angle and position precision of 0.001° and 5 µm, respectively. This PDMS/PC stamp is used to uniformly dry-adhere and subsequently break-off the target membrane from the tether. The membrane is then flipped by introducing another PDMS stamp and placed on a hydrogen silsesquioxane (HSQ) - coated carrier wafer (see Figure S4 in[32]). Next, the structure is annealed at 600 °C to allow the HSQ to transition into a denser $SiO_x$ film.[38] The $SiO_x$ film has a low thermal expansion ratio,[39] low optical loss,[13] is photochemically inert, and is compatible with most lithography schemes. The success of this work flow depends on the adhesion differences between PC, PDMS, and HSQ layers. This guarantees a transfer process that is highly deterministic, reproducible, and agnostic to carrier wafer and surface pattern choices. The overall height variation across the whole membrane is determined to be $\sigma \leq 10$nm (see Figure S5e in[32]). In this work, we bond diamond membranes to fused silica and thermal oxide wafers with pre-defined trenches, generating locally suspended regions. Suspension allows us to control the chemical termination of both surfaces and reduce HSQ-related fluorescence for photoluminescence (PL) characterizations of the embedded point defects(see section 1.8 and 2.4 in[32] for details). Finally, by flipping the membrane, we are able to fully etch-away the $He^+$-damaged, graphitized diamond layer with chlorine-based ICP. This ICP etching eliminates the undesired fluorescence and built-in strain caused by crystallographic imperfections and lattice mismatch.[27,32] The right part of Figure 1f shows a 20 h grown diamond membrane on a fused silica wafer with 100 nm final thickness. This etching step also serves to precisely tune the thickness of the final diamond structure(see section 1.4 in[32]), with demonstrated thicknesses as low as 50 nm.[13] As such, this membrane synthesis approach provides a clear, high yield, scalable, and easy path forward for generating multi-field relevant diamond membranes.

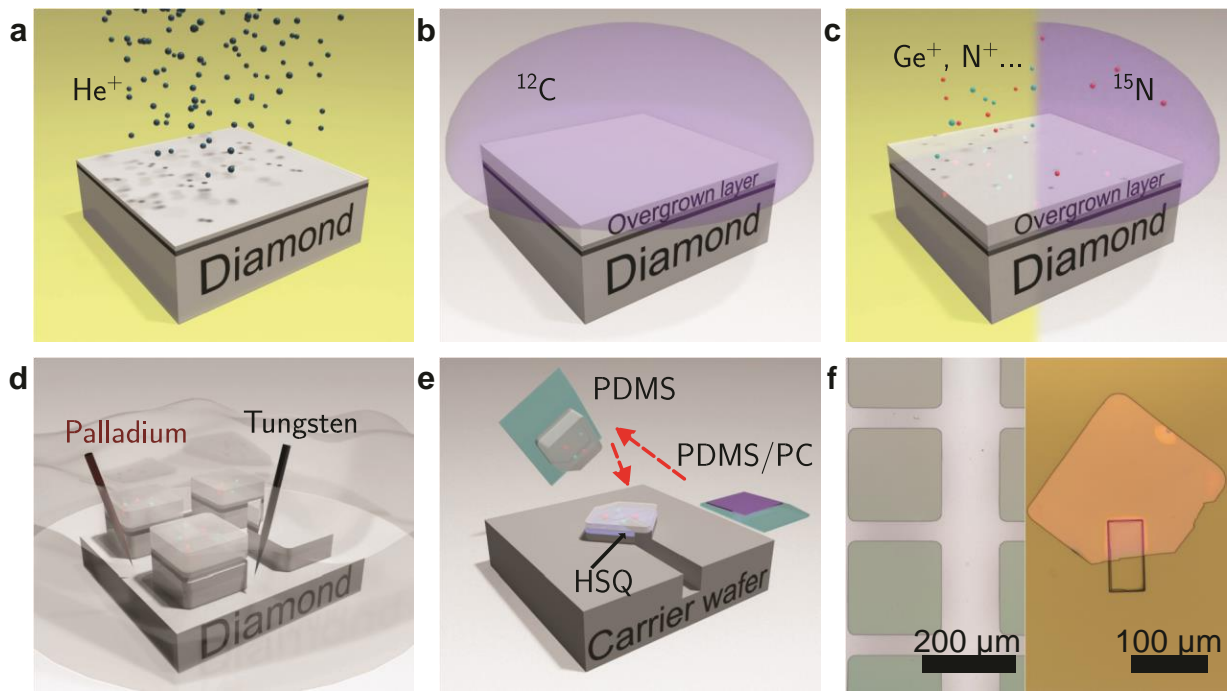

**Figure 1**: Schematics of the diamond membrane fabrication process. (a) He$^+$ implantation with subsequent annealing to form the membrane (light gray on the top) and the graphitized layer (dark grey underneath). (b) Isotopically ($^{12}$C) purified PE-CVD diamond overgrowth. (c) Color center incorporation via either ion implantation (left) or in-situ doping (right). Red dots: N$^+$. Blue dots: Ge$^+$. Other implanted species (Si$^+$, Sn$^+$) are not shown. (d) Diamond membrane undercut via EC etching. The undercut takes place in DI water, with palladium anode (red) and tungsten cathode (dark grey) aiming at the target membrane. (e) Membrane transfer and back etching. The membrane is picked up by the PDMS/PC stamp (green/purple), flipped onto another PDMS stamp (green), and bonded to the carrier wafer by HSQ resist (the blue layer sandwiched in between). (f) Microscope images of patterned overgrown membranes (left) and a transferred and multi-step etched membrane on a fused silica wafer (right). The green squares on the left are patterned membranes with underneath graphitized layer.

## Surface and Material Quality

The utility of diamond membranes in QIS demands exceptional surface and material quality. Figure 2 shows the atomic force microscopy (AFM) characterizations of the membrane backside (a-b) and topside (c) topology following all processing. We note that for both growth conditions (500 °C for 20 h and 700 °C for 40 h), the resulting growth surface (i.e., the backside after dry transfer) is smooth, showing distinct step-flow growth striations,[40] with a roughness ($R_q \leq 0.31$ nm) lower than the diamond lattice constant (0.357 nm). Similarly, the topside, although initially rough due to the intrinsic straggle of the He$^+$ implantation process (see Figure S1e in[32]), reaches $R_q$ of $\leq 0.3$ nm following ICP etching,[41] as shown in Figure 2c. The realization of atomically flat surfaces is critical for effective surface termination and coherence protection of near-surface color centers.[33]

Additionally, the diamond crystal quality was investigated via room temperature Raman spectroscopy as shown in Figure 2d. To benchmark the spectra, we compare our values to a surface strain-released[36] reference diamond (single-crystal, electronic (EL) grade) from Element Six, with a Raman linewidth of 1.570(1) cm$^{-1}$. The ≈185 nm overgrowth membrane (20 h growth in 500 °C, back-etched down to 100 nm) presents a Raman linewidth of 1.779(5) cm$^{-1}$, slightly larger than the reference value. Remarkably, the isotopically purified, ≈370 nm overgrowth membrane (40 h growth in 500 °C, back-etched down to 110 nm) presents a linewidth of 1.375(2) cm$^{-1}$ (1.486(14) cm$^{-1}$ for the ≈250 nm isotopically purified overgrowth at 700 °C, see Figure S2b in[32]), significantly lower than anything reported previously,[28] even compared to pristine natural isotopic abundance bulk diamond. The ultra-narrow Raman peak indicates the crystal is free of impurities and defects. The up-shifting and narrowing, in comparison with the bulk spectra, is consistent with the change of the Raman transition brought on via isotopic purification of a high-quality diamond structure.[29] As such, the Raman spectra indicate that these diamond membranes are of ideal crystal quality to host color centers.

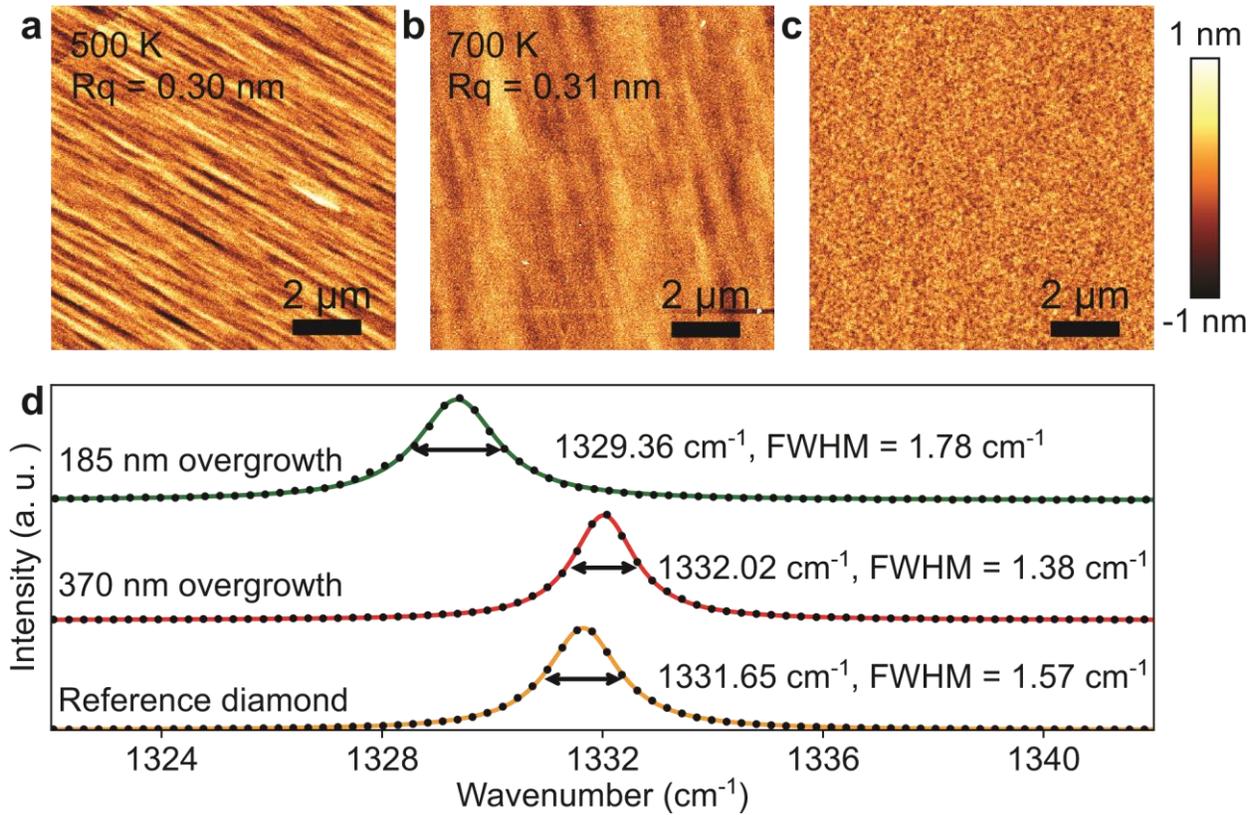

**Figure 2**: Surface morphology and crystal quality of overgrown membranes. (a-b) AFM images of overgrowth patterns (back side of membranes after the transfer) at different heating plate temperatures. (c) An AFM image on the front side of the membrane after multi-step etching. (d) Room temperature Raman spectroscopy of diamond membranes and the reference diamond substrate. Green: The ≈185 nm (20 h) overgrowth membrane back-etched down to 100 nm. Red: The ≈370 nm (40 h) overgrowth, isotopically purified membrane back-etched down to 110 nm. Yellow: A surface strain-released, EL grade single crystal diamond used as the reference. The surface strain is polish-induced, and is removed by ICP etching and subsequent annealing prior to the Raman spectroscopy(see section 2.2 in[32]).

## Implanted Group IV Color Centers

In order to investigate the coherence of color centers in the membranes, we performed detailed measurements of the optical (spin) coherence of GeV⁻ (NV⁻). We mounted the 110 nm thick (≈370 nm overgrowth), $^{12}$C-membrane on a trenched thermal oxide wafer with a vapor HF-based HSQ removal (see section 2.4 in[32] for details), as shown in Figure 3a. This HSQ removal is optional, and the additional HSQ fluorescence has minimal impact on device performance for GeV⁻ resonant experiments. GeV⁻ centers inside the sample were subsequently characterized at 5.4 K. This particular membrane was also implanted with nitrogen (see further in the text), silicon, and tin (see section 1.5 in[32]).

A PL map of the GeV⁻ centers over the suspended area is shown in Figure 3b. The GeV⁻ conversion efficiency was determined to be 6.5(4)% (see section 3.2 in[32]). Typical zero phonon line (ZPL) peaks of GeV⁻ centers lie between values reported in bulk diamond,[42,43] indicating a homogeneous crystal environment (see Figure S6a in[32]). The span of the ZPL peaks across different centers come from the nature of the implantation.[44] The inhomogeneous broadening of the ZPL positions are comparable with those measured in bulk diamond under identical implantation and annealing conditions. While strain generated from the membrane mounting, thermal expansion ratio mismatch with the carrier wafer, and HSQ annealing may potentially contribute to energetic variation, our measurements indicate the crystal environments between the diamond membrane and bulk diamond are highly similar (see section 3.5 in[32]). Both off- and on-resonance autocorrelation measurements were performed on the GeV⁻ centers (see section 3.3 in[32]), confirming anti-bunching features associated with single photon emitters with $g^{(2)}(0) = 0.17(2)$ and 0.19(5), respectively. Additionally, the transition linewidths of the GeV⁻ centers are relatively low, with single scan lines as narrow as 70(1) MHz. Furthermore, this optical transition is stable, offering a 2h average spectral broadening as low as 124.8(2) MHz, as seen in Figure 3c. This broadening is still less than a five-fold increase over the intrinsic linewidth.[42]

For completeness, we also analyzed the statistical distribution of single-scan linewidths and time-averaged broadening. Out of 38 centers observed with off-resonant excitation, we were able to resolve the C transition in 29 centers with resonant scanning. Additionally, in 7 (1) centers, we observed the C transition switches between 2 (4) stable peaks (see section 3.4 in[32] for the 2 peaks pattern). The origin of this switching between two stable energies, distinct from random spectral diffusion, is the subject of ongoing study. Statistics of the other 21 single-peak GeV⁻ centers yield a median single (2.5 min average) scan linewidth of 95 MHz (231 MHz), as shown in Figure 3d. Recent cavity quantum-electrodynamics measurements[45] have realized a spin-photon coupling rate of $g = 2\pi \times 5.6$ GHz, significantly larger than the observed spectral broadening in our centers. Therefore, the membrane-hosted GeV⁻ centers are sufficiently optically coherent for advanced applications in quantum networking and entanglement generation. These diamond membranes are also viable hosts for SiV⁻ and SnV⁻ centers (see section 3.3 in[32]).

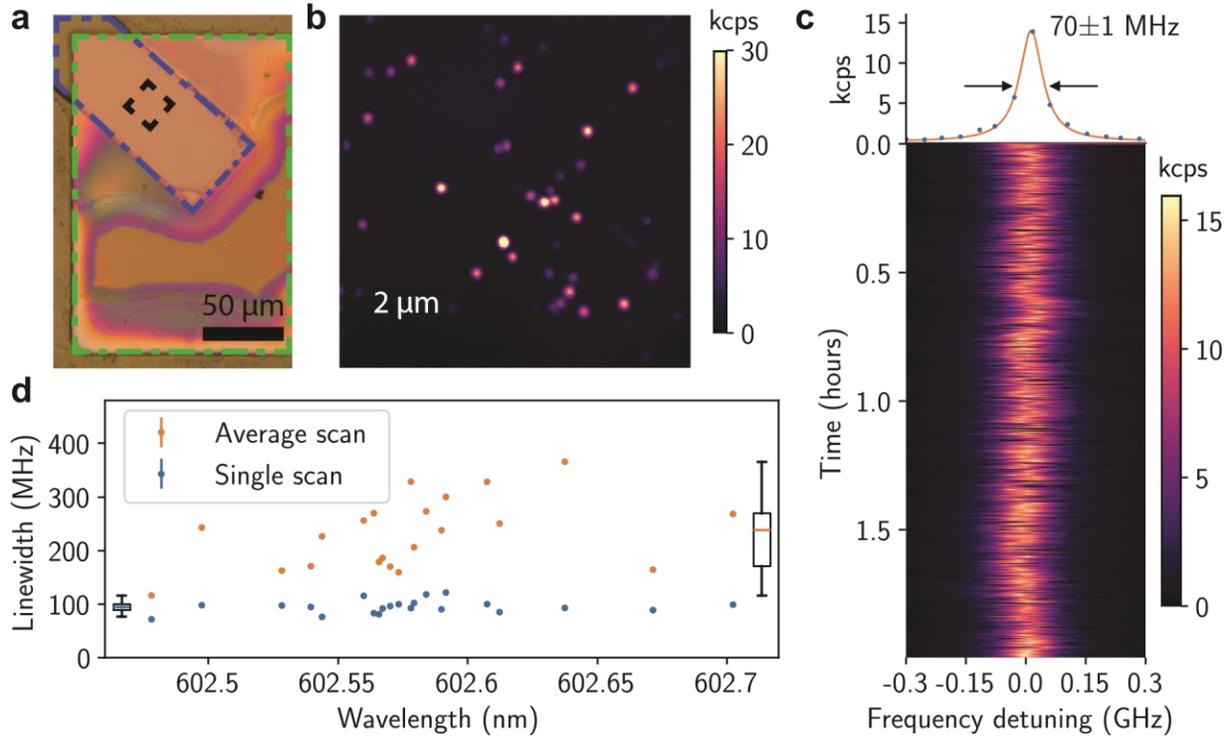

**Figure 3**: Optical characterization of GeV⁻ centers at 5.4 K. (a) A microscope image of a diamond membrane (dashed green) containing implanted GeV⁻ centers. HSQ in the trench (dashed blue) is completely removed by vapor HF. The rainbow color on the membrane indicates excess HF undercut (see section 2.4 in[32]). The PL measurement is performed in the dashed black region. (b) A PL map of the implanted GeV⁻ centers. The signal-to-background ratio for most centers are between 5 to 30. (c) Single PLE scan and time-averaged broadening of a GeV center. The data sampling time per point is 100 ms, and the counts are normalized to kilo-counts-per-second (kcps). (d) Statistics of GeV⁻ optical linewidths measured via single-scan (blue) and average broadening (orange). The inset box plot on the left (right) indicates the median single (average) linewidth of 95 MHz (231 MHz).

**Embedded NV⁻ Centers**

Next, we investigated the spin coherence of NV⁻ centers in the membranes using a home-built room temperature PL microscope (see section 4.1 in[32]). While the Zeeman splitting frequencies of group IV diamond color centers are one to two orders of magnitude larger than those of the NV⁻ center,[2,46,47] NV⁻ coherence measurements reveal magnetic noise levels which are also relevant to group IV coherent control.[33,48–50] All NV⁻ center measurements presented herein were performed with a 15 gauss static magnetic field applied at 10° angle to the [111] crystal axis. We measured NV⁻ centers along all four possible crystal orientations, determined by different transition frequencies.

Figure 4a-b show representative free induction decay and spin echo decay curves on a single long-lived NV⁻ spin, with the fitted $T_2^*$ and $T_2$ coherence times. The oscillations in the first 100 μs of Figure 4b arise from aliasing of electron spin echo envelope modulation (a finer trace to demonstrate the origin of this is presented in the Figure S7c of[32]). Together, these measurements demonstrate that membrane fabrication does not preclude the formation of highly coherent spin qubits. Figure 4c presents a scatter plot of the $T_2^*$ and $T_2$ times, showing a spread of 4.3(3) μs to 149(7) μs and 8(2) μs to 400(100) μs, with most times (median 50%) falling above 10 μs and 30 μs, respectively.

A common source of magnetic noise for NV⁻ centers is unconverted nitrogen atoms (P₁ centers). Taking into account nitrogen originating from both implantation and background in-grown sources, we estimate a density [$N$] of ≤ 1ppb (see section 4.5 in[32]). At these concentrations, P₁s should not contribute heavily to the decoherence.[51] Comparing the background nitrogen concentration and the observed areal NV⁻ density versus the [$N$] implantation dose of $2 \times 10^8 \, cm^{-2}$, we expect that many of our observed NV⁻ formed from in-grown nitrogen and vacancies introduced during ion implantation and not exclusively from the implanted nitrogen. Thus, the observed NV⁻ are likely distributed throughout the thickness of the membrane, with some residing within ≤15 nm of both surfaces, where previous work demonstrated marked decoherence from surface noise.[49] This ambiguity can be avoided by implanting $^{15}$N in future iterations of this experiment. However, statistically, the surface proximity distribution of the NV$^-$ alone cannot fully account for the large number of NV centers with $T_2 \leq 100$ µs. The multi-species implantation process is known to introduce crystal damage throughout the ion path, which can create spin-full vacancy complexes that are not mobilized, nor annihilated during the annealing process.[50] It is likely that the resulting inhomogeneity of the bulk spin bath is the main factor limiting NV$^-$ coherence times. Nonetheless, the spin echo coherence time ($T_2$ up to 400 µs) is competitive with near bulk-like properties, and the free induction decay ($T_2^*$ up to 150 µs) outperforms commercially available bulk material due to the $^{12}$C purification. Therefore, the coherence times presented herein are fully compatible with applications in quantum sensing and hybrid quantum systems.[48,52,53]

Additionally, we utilize NV$^-$ optically detected magnetic resonance (ODMR) measurements to characterize strain in the diamond membrane, which may be introduced by the fabrication. ODMR measurements at zero static field reveal splittings between upper and lower NV$^-$ transitions that are generally larger than shifts in the zero field splitting. We thus attribute these effects mainly to local electric field variations (see section 4.3 in[32]).[54] If we take the 1 MHz observed shifts as an upper bound on strain, we find a maximum strain of $10^{-4}$.[55]

This membrane fabrication technique is also highly amenable to in-situ $\delta$-doping of $^{15}$N during overgrowth.[1,32] $\delta$-doping allows deterministic incorporation of dopants (N, Ge, Si, etc.) during membrane overgrowth while providing a valuable distinction from the intrinsic, isotopically naturally abundant defects (i.e., isotopically incorporating $^{15}$N during growth to distinguish from the $^{14}$N overgrowth background). As a proof of concept we introduced 2 nm-thick area of $^{15}$N doping ≈36 nm from the backside of a 110 nm thick diamond membrane (≈250 nm overgrowth at 700 °C). Figure 4d shows a PL map of NV centers in such a sample. The $^{15}$NV$^-$ centers are labeled in teal circles, while background $^{14}$NV$^-$ are in white rectangles, with representative hyperfine-resolved ODMR spectra presented to the right of the figure. We observed a 7 : 11 ratio of [$^{15}$NV$^-$] : [$^{14}$NV$^-$] (SIMS characterized [$^{15}$NV$^-$] of 30.8(57) ppb, see section 4.5 in[32]). This is in a good alignment with what was observed for the implantation-synthesised NV⁻ centers, showing a consistent background $^{14}$NV⁻ density throughout the membrane from the overgrowth process. A rigorous quantitative comparison of optimized implanted and in-gown defects as they relate to the membrane surfaces proximity is left for subsequent studies to explore.

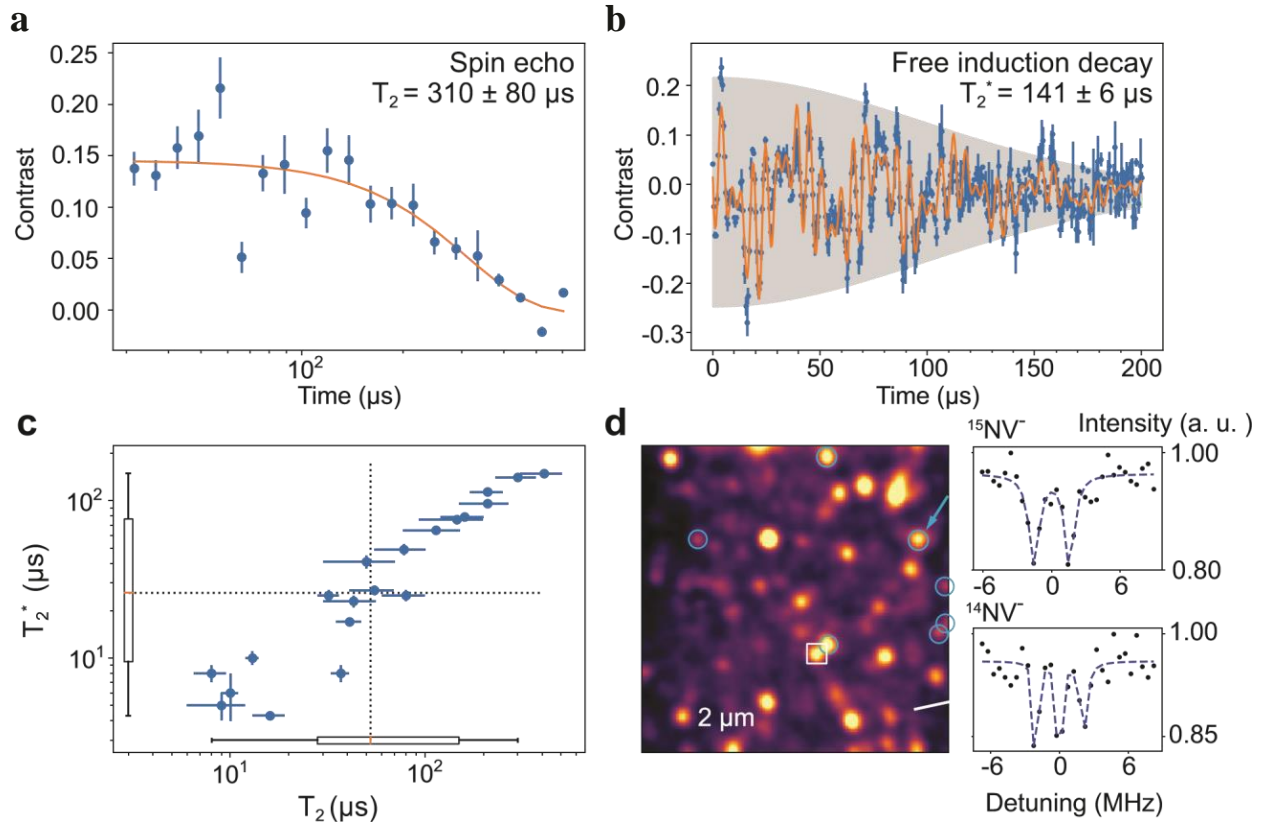

**Figure 4**: Optical characterization of embedded NV⁻ centers at room temperature. (a-b) Representative free induction decay and spin echo decay curves on a single long-lived NV⁻ spin, accompanied by the $T_2^*$ and $T_2$ coherence times. A detailed description of data analysis is provided in the Supporting Information. The oscillations in the first 100 µs of (b) arise from aliasing of electron spin echo envelope modulation. (c) A scatter plot of $T_2^*$ and $T_2$ times for the 20 measured NV⁻ centers. Inset box plots denote median values of 26 µs and 52.5 µs (dashed lines) and lower-quartile values of 9.5 µs and 28 µs. Error bars are fit errors. (d) NV⁻ PL map of a δ-doped membrane with $^{15}$NV⁻ centers (teal circles) and $^{14}$NV⁻ centers (white squares) labeled. At right, pulsed-ODMR spectra of the indicated NV⁻ centers.

# Conclusion

We have created uniform, tunable, transferable, large-scale diamond membranes with material and surface quality comparable to and even exceeding bulk diamond. Additionally, we have demonstrated that color centers within the membranes have sufficient optical and spin coherence for a broad range of applications in QIS. These diamond membranes can serve as a quantum material platform for a wide range of current and future research directions. The sub-wavelength thickness of the diamond membranes expands capabilities for nanophotonic integration.[13] Additionally, this platform eases the fabrication of strain engineering[56] and Stark tuning,[57] which also provides an opportunity to study behavior of color centers under extreme environments such as strong strain or electrical fields. Moreover, the versatility afforded by the arbitrary carrier wafer platform is vital for integrating diamond-based quantum systems to phononic,[12,58] superconducting, and magnonic[59] systems for quantum transduction in a hybrid geometry.[60] Furthermore, $NV^-$-based quantum sensing will benefit from the ease of integration with a number of interfaces, ranging from type-II superconductors to living cells.[15,16] The isotopically purified, highly crystalline diamond membranes are also prime candidates for heat dissipation device studies[61] and high quality engineered nanodiamonds synthesis.[62,63]

# Acknowledgement

This work was primarily supported by the U.S. Department of Energy, Office of Science, Basic Energy Sciences, Materials Sciences and Engineering Division with support from the U.S. Department of Energy, Office of Science, National Quantum Information Science Research Centers. A. H. was partially supported by the University of Chicago Materials Research Science and Engineering Center, which is funded by the National Science Foundation under award number DMR2011854. This work made use of the Pritzker Nanofabrication Facility part of the Pritzker School of Molecular Engineering at the University of Chicago, which receives support from Soft and Hybrid Nanotechnology Experimental (SHyNE) Resource (NSF ECCS-2025633), a node of the National Science Foundation's National Nanotechnology Coordinated Infrastructure. This work also made use of the shared facilities at the University of Chicago Materials Research Science and Engineering Center, supported by the National Science Foundation under award number DMR-2011854. Funding was provided by the Boeing company and the University of Chicago Joint Task Force Initiative. J. K. and A. B. acknowledge support from the NSF Graduate Research Fellowship under grant no. DGE-1746045. The authors thank A. S. Greenspon for experimental help.

# Supporting Information: Tunable and Transferable Diamond Membranes for Integrated Quantum Technologies

Xinghan Guo[1], Nazar Delegan[1,2], Jonathan C. Karsch[1], Zixi Li[1],
Tianle Liu[3], Robert Shreiner[3], Amy Butcher[1], David D. Awschalom[1,2,3],
F. Joseph Heremans[1,2], Alexander A. High[1,2]*

[1]*Pritzker School of Molecular Engineering, University of Chicago, Chicago, IL 60637, USA*
[2]*Center for Molecular Engineering and Materials Science Division,*
*Argonne National Laboratory, Lemont, IL 60439, USA*
[3]*Department of Physics, University of Chicago, Chicago, IL 60637, USA*
*E-mail: ahigh@uchicago.edu

## 1 Diamond membrane fabrication procedure

### 1.1 Membrane synthesis via the graphitized underlayer formation

Standard 3 mm by 3 mm by 0.25 mm single crystal, optical grade diamond substrates (Element Six, ≤1 ppm [N]) are used for the membrane synthesis. These are initially fine-polished to a surface Rq of ≤0.3 nm (Syntek LLC.) as to minimize morphological inconsistencies (see Figure S1 (a)). Next, the samples are implanted with $^4$He$^+$ ions (CuttingEdge Ions LLC.) at 150 keV to create a graphitized layer at ≈410 nm depth. This is done at an incidence angle of 7° to avoid ion channeling. The dose is set to $5 \times 10^{16}$ cm$^{-2}$, which is closely above the graphitization threshold as to minimize crystal damage (see section 1.5). In this work, a threestep annealing process was employed after the implantation: a 400 °C soak for 8 h, followed by a 800 °C soak for 8 h, completed with a 1200 °C anneal for 2 h.[1] This was done in a forming gas environment (Ar:H$_2$ of 96 : 4). The implantation and annealing had no negative impact on the surface roughness (see Figure S1 (b)). The phase transition of the carbon bonds during membrane formation is studied via room temperature Raman spectroscopy (see section 2.2).

### 1.2 PE-CVD diamond overgrowth and $\delta$-doping of $^{15}$N

Homoepitaxial PE-CVD of diamond is performed in a custom configured *SEKI DIAMOND SDS6350* system. Before loading into the growth chamber, the diamond substrates are chemically tri-acid cleaned (see section 1.7). The diamond sample is placed on a grooved molybdenum substrate to keep it flat with the rest of the plasma exposed surface, allowing for a uniform growth rate across the sample, while minimizing edge effects. The chamber is pumped down to a $2 \times 10^{-8}$ Torr to $5 \times 10^{-8}$ Torr base pressure in order to reduce background contamination. Thereafter, high purity $H_2$ (99.999999% chemical purity) is introduced into the chamber, with the gas flow ratios constant and pumping rate monitored as to maintain a constant process pressure of 25 Torr. Throughout the process the microwave power is maintained at 900 W (11.5 Wmm$^{-2}$). Before introducing the carbon precursor, the substrate is heated to a 500 °C or 700 °C setpoint which translates to a pyrometer assessed plasma temperature of 540(10) °C or 740(10) °C, respectively. The $H_2$-only plasma conditions are maintained for 20 min (500 °C) and 1 h (700 °C) as to etch away any residual surface carbonaceous contaminants – the approximated sp$^3$ etch depth is ≤1 nm.[2] Thereafter, $^{12}CH_4$ (99.99999% chemical purity, 99.99 at.% isotopic purity) is introduced as the carbon precursor, this injection is substantiated in-situ via increase of the pyrometer detected substrate surface temperature to 765(10) °C (580(10) °C for the 500 °C set-point grown substrate). The methane-to-hydrogen ratio is maintained constant at 0.05% ($H_2$:$^{12}CH_4$ = 400 sccm : 0.2 sccm) as to ensure step-flow growth.[3,4] Growth rates for the 700 °C and 500 °C membranes were determined to be 6.2(4) nmh$^{-1}$ and 9.3(8) nmh$^{-1}$, respectively (see section 2.3). The reduction in growth rate is explained by an increased surface mobility and desorption of precursor adatoms[5] and likely due to a non-linear reduction of nucleation sites.

In-situ nitrogen doping is accomplished by introducing 0.06 sccm of $^{15}N_2$ gas (99.99% chemical purity, 99.9 at.% isotopic purity) for 2 min during the growth phase. This creates a 36nm depth localized layer with a SIMS quantified [$^{15}NV^-$] density of 30.8(57) ppb (see section 4.5). Thereafter, a cap layer is overgrown to encapsulate the $\delta$-doped layer as deep in the membrane as desired. Subsequent irradiation and annealing allows the conversion of this layer into ≈0.9 ppb of [$^{15}NV^-$], resulting in an average color center creation efficiency of ≤2.5%, in agreement with previous studies (see section 4.5).[3]

## 1.3 Membrane patterning and EC etching

Atomic layer deposition (ALD) is used to deposit 25 nm of $Al_2O_3$ on the overgrown membrane samples. This oxide layer serves as selective hard mask for subsequent ICP etching. Thereafter, a photoresist layer (AZ MiR 703) is applied to the sample, followed by a lithography exposure (Heidelberg MLA150, 375 nm laser, with a 150 mJcm$^{-2}$ dose). A total of 56 square-shaped membranes are patterned on a single substrate. Next, the sample is developed (AZ 300 MIF) and two-step etched in a chlorine-based ICP chamber (Plasma - Therm, $Al_2O_3$ etched with $BCl_3$ chemistry first, with diamond subsequently etched in $O_2$ chemistry, see Table S2). This step exposes the underlying graphitic layer. After etching, the sample is ultrasonicated in 80 °C NMethyl-2-pyrrolidone (NMP) solution and placed in a buffered oxide etch (BOE) solution to remove excess resist and oxide layers.

For the EC etching step, the sample is held by double side tapes at the bottom of a DI waterfilled petri dish which is mounted on a probe station (Signatone S1160). Unlike previous studies which required customized platinum tips,[6] we use commercially available tungsten tips (Signatone SE-T) as the cathode and palladium tips (Signetone SE-P) as the anode (see Figure S3 (b)). These tips are mounted to 3-axis micropositioners (Signatone S926), which are connected to a DC power supply (B&K Precision 1787B). Typically, a voltage between 15 V and 36 V is applied to undercut the membranes, with larger undercut areas requiring larger potentials and longer time. The EC etch is visually tracked, with the voltage cut at the moment that the desired undercut tether has been achieved. This process, while relatively simple, requires constant supervision as the final tether feature size is typically less than 50µm in any direction. Practically speaking, at the typical voltages quoted above, this translates to an etching time of ≈30 min per membrane (see Figure S3 (c)). This processing step is a great candidate for future automation.

## 1.4 Membrane transfer and post-processing

Prior to the membrane dry transfer, the intended carrier wafers or devices are prepared with pre-defined structures and features (see section 1.8). In this work, we use 7 mm by 7 mm fused silica ($SiO_2$, JGS2 grade) and thermal oxide (280 nm $SiO_2$_Si) wafers. Alignment markers and 5 µm deep trenches are patterned and etched into substrates to suspend part of the transferred membrane (see Figure S5 (a)). In this way, both surfaces of the membranes are exposed, enabling double-sided surface termination via wet-chemical or dry processes with acid cleaning and oxygen annealing used in this work. As for the dry-transfer itself, we start by mounting the diamond substrate on a transfer station rotational chuck (Signatone S1160). Parallel to this, we prepare a PDMS/PC stamp on a glass slide and mount it into a computer-aided micropositioner (Signatone CAP - 946). Next, we slowly bring the stamp down and allow it to fully adhere to the target membrane. With a quick movement of the now-adhered stamp, we break the membrane from the tether. Then, we replace the substrate on the chuck with a second PDMS stamp on a glass slide. The detached membrane can be transferred to this second stamp by slowly bringing the PDMS/PC stamp into contact with the "stickier" PDMS-only stamp and lifting the PDMS/PC stamp afterwards. Finally, with the membrane fully adhered to the PDMS-only stamp, we use the micropositioner to mount the membrane to an HSQ-covered carrier wafer on the chuck. After careful alignment, we gradually bring the stamp down until the membrane is fully attached to a freshly spin-coated HSQ layer (14% HSQ in methyl isobutyl ketone (MIBK) solution, DisChem Inc.), and then slowly release the stamp. The membrane is left on the carrier wafer due to stronger adhesion from the HSQ polymer and the residual MIBK solvents (see Figure S4).

After membrane transfer, the carrying wafer is annealed at 600 °C for 8 h in Ar environment. The HSQ adhesion layer fully collapses during the annealing and forms a nearly strain-free ≈250 nm thick layer. This thickness can be tuned by using different concentration of HSQ solvents or by varying the spinning parameters. In the case of partially suspended membranes, we applied a vapor HF treatment (Memsstar ORBIS ALPHA) to fully remove the HSQ in the trench (see section 2.4). Additionally, in order to etch the original damaged underlayer (now on the top face of the membrane), improve surface morphology, and tune the final thickness, we submit the mounted membrane to a three-step ICP process (see Table S2).[1] We start with the "Ar/$Cl_2$" recipe (2 min to 5 min, depending on the target thickness) to polish the surface, then we use the "$O_2$/$Cl_2$" recipe (30 s to 90 s) to remove most of the chlorine-based compounds on the diamond surface. Finally, we apply a 30 s "$O_2$" etch to remove the residual chlorine compounds and correctly terminate the surface. In order to avoid chemical contamination, etching steps are separated by multiple pump-purge cycles. AFM characterizations are added post-etching to ensure no micromasking occured during ICP processes.[7] The membrane thickness is monitored via either a profilometer (Bruker DektakXT) or an ellipsometer (Horiba UVISEL2) (see section 2.3).

## 1.5 Additional information on ion implantation

Two distinct ion implantation runs were performed in this work: $^4$He$^+$ ion implantation for the creation of the graphitized underlayer and $^{14}$N$^+$, $^{28}$Si$^+$, $^{74}$Ge$^+$, and $^{120}$Sn$^+$ ion implantation for the creation of color centers. Both implantation processes were performed at CuttingEdge Ions, LLC.

To achieve the formation of the graphitized underlayer, a low energy $^4$He$^+$-ion implantation into the diamond substrates was done as depicted in Figure 1 (a). This step forms a depthlocalized graphitized underlayer ≈410 nm deep via a damage-induced phase transition from sp$^3$ to sp$^2$. The implantation energy was kept at 150keV to minimize crystal damage along the $^4$He$^+$ trajectory. The dose was set to $5 \times 10^{16}$ cm$^{-2}$, which is above the graphitization threshold, but minimized to avoid unnecessary crystal damage. Separate implantation tests showed that lower doses of ≤$2 \times 10^{16}$ cm$^{-2}$ were insufficient to generate a reliable graphitized layer, with the diamond substrate remaining transparent. Therefore, the desired damage threshold was established to be between $2 \times 10^{16}$ cm$^{-2}$ to $5 \times 10^{16}$ cm$^{-2}$.

We use the Stopping and Range of Ions in Matter software (SRIM)[8] to estimate the ion implantation depth and straggle. For the simulations, the density of carbon is set to 3.51 gcm$^{-3}$, and the incidence angle is 7° to avoid ion channeling. For color center incorporation, we simultaneously implanted $^{14}$N$^+$, $^{28}$Si$^+$, $^{74}$Ge$^+$ and $^{120}$Sn$^+$ at a dose of $2 \times 10^8$ cm$^{-2}$ in order to achieve an individual-level color center density. Details about the implantation parameters and simulated depths and straggles are shown in Table S1.

All of the implantation steps were followed with the three-step 1200 °C anneal[1] (see section 1.1) to form the graphitized layer and color centers, in the case of the He$^+$ and color center implantation, respectively.

## 1.6 Electron irradiation

For the synthesis of delta-doped color centers, an electron radiation of $10 \times 10^{18}$ cm$^{-2}$ at 2 MeV at the low energy accelerator facility (LEAF) at Argonne National Laboratory was performed to form vacancies throughout the diamond crystal. This was followed by a 6 h vacancy mobilization anneal at 850 °C in forming gas (96 : 4 of Ar:H$_2$).

## 1.7 Surface cleaning and termination

In order to remove contamination and passivate the surface, acid-related surface cleanings and terminations are applied multiple times during membrane synthesis. Firstly, diamond substrates receive a tri-acid cleaning (1 : 1 : 1 nitric, sulfuric, and perchloric acid) at 200 °C for 2 h right before the overgrowth to eliminate surface contamination[9]. Secondly, the diamond membrane template receives a di-acid cleaning (1 : 1 nitric and sulfuric acid) at 225 °C for 2 h after every transfer step to prevent accumulated contamination induced by EC etching or PDMS/PC stamping. Finally, before optical measurements, membrane samples receive another di-acid cleaning and an optional oxygen termination, depending on the type of color centers being characterized. For group IV centers, we only apply the di-acid clean procedure. For NV center measurements, an additional oxygen termination is applied following di-acid cleaning, which includes a 4 h annealing at 400 °C in O$_2$ atmosphere, with 5 min rinse in freshly prepared Piranha solution (3 : 1 sulfuric acid and hydrogen peroxide) before and after the annealing. The oxygen termination is intended for optimizing the NV$^0$ to NV$^-$ conversion and stability[10]. The incorporation of higher temperature oxygen annealing while avoiding surface etching is an ongoing study and beyond the scope of this work.

## 1.8 Carrier wafer preparation

In this work, we dice thermal oxide and fused silica wafers to 7 mm by 7 mm squares (Disco DAD3240) as carrier wafers. Some wafers receive ≈5 µm trench fabrication on the surface to allow membranes to receive double-side surface treatment and obtain low background fluorescence during optical measurements.

In order to etch the trenches, we deposit a ≈6 µm photoresist layer (AZ 4620) as the etching mask. The spinning speed is set to 500 rpm for 10 s and then 4000 rpm for 50 s, with a ramping rate of 100 rpms$^{-1}$ and 2000 rpms$^{-1}$, respectively. After resist spinning and pre-baking, the sample is lithographically exposed (Heidelberg MLA150, 405 nm laser with 320 mJcm$^{-2}$ dosage), followed by development in AZ 400K 1:4 solution. We use fluorine-based ICP (Plasma - Therm) for SiO$_2$ and Si etching. After etching, samples are ultrasonicated in 80 °C N-Methyl2-pyrrolidone (NMP) solution to remove the resist.

## 1.9 Additional discussions on diamond membrane fabrication

The membrane creation process highlighted herein advances scalable diamond quantum research in several important ways. While the process requires initial diamond substrates with sufficient crystallinity and smoothness, it is agnostic to moderate impurity and defect concentrations within the substrate. This eliminates the need for highly purified bulk diamond substrates, which are costly and scarce. Additionally, while bulk diamond crystals are commercially grown and available for purchase in a limited range of dimensions, the size and the shape of the membranes are fully defined and can be tailored to specific applications, with the maximum size only limited by the substrate dimensions. While our choice of membrane size (200 µm × 200 µm) is well motivated by scientific applications, such delaminations over millimeter size have been achieved[11], and there are no technical barriers to realizing larger dimensions with our technique.

# 2 Material characterization

## 2.1 AFM characterization

We extensively apply AFM measurements (Bruker Dimension Icon) to monitor the surface morphology throughout the synthesis steps. The scanning ranges and resolutions in both directions are set to 10 µm and 512 lines (except for Figure S1 (f), 5 µm with 256 lines) to observe the surface morphology and growth- or etching-related features.

The surface maps of overgrown and back-etched diamond membranes are shown in Figure 2 (a-c) of the main text, with maps of other key steps shown in Figure S1. Since the as-purchased diamond substrates are only roughly polished ($Ra \leq 30$ nm), we applied another fine-polish (Syntek LLC.) step to reduce the roughness to $Rq = 0.30$ nm ($Ra = 0.20$ nm), as shown in Figure S1 (a). The final surface miscut, relative to the crystallographic (001) axis, is specified to ≤3°. Unfortunately, while this value is relatively consistent within a single substrate batch, it can vary between substrate lots, affecting the effective implantation angle and overgrowth characteristics[4].

The He$^+$-graphitized and subsequently annealed substrate is shown in Figure S1 (b), with the same level of surface morphology ($Rq = 0.27$ nm). Figure S1 (c) exhibits the 40 h overgrown diamond surface with a pre-growth 60 min hydrogen plasma etch. The inconsistency of surface roughness under the same duration of hydrogen plasma etch (see Figure 2 (b)) may come from the residual surface strain from fine polishing. A surface strain release step[1] after the fine-polish should improve this in future iterations of the fabrication process. Another 40 h growth without proper surface preparation is shown in Figure S1 (d). Residual contamination on the diamond surface generates growth defects (such as pits or pyramids) along the process. Figure S1 (e) shows the front side of the diamond membrane after the dry transfer. The excess roughness ($Rq = 1.17$ nm) comes from the straggle of the He$^+$ implantation and can be reduced by subsequent "Ar/Cl$_2$" etching. It is crucial to apply "Ar/Cl$_2$" to prevent roughness deterioration. Figure S1

(f) shows the front side of the membrane if the O₂ etching is applied directly, where the surface deteriorates noticeably due to the preferential etching nature of an O₂-based plasma on crystallographic defects and polish-induced damage.

## 2.2 Raman characterization

Throughout the process, we applied Raman spectroscopy (Horiba Scientific LabRAM HR Evolution) to investigate both the phase of the carbon during membrane formation and the diamond crystal quality. A 633 nm excitation laser is used instead of the more popular 532 nm to avoid broad fluorescence from NV⁻ centers in the substrates. We use the finest grating (1800 grmm$^{-1}$) for data collection, with a quoted resolution of ≈0.3 cm$^{-1}$. The resulting peak position uncertainty is caused by this instrument resolution, while the linewidth uncertainty comes from the fittings. The pinhole of the Raman confocal microscope is set to either 30% or 50% to reduce artificial broadening.

To effectively evaluate the crystal quality of our overgrown membranes, we used a reference diamond for the Raman measurements. In order to achieve the cleanest Raman signature, we chose an electronic grade single crystal diamond from Element Six. Furthermore, we ICPetched the top ≈5 µm of the reference sample, and subsequently annealed the substrate in order to remove the residual strain induced by surface polish. The etching procedure is described in[12], with our exact process flow detailed in Table S2.

Figure S2 (a) shows the comparison of implanted diamond substrates pre- and post-annealing. Prior to annealing, only a weak diamond peak from the back substrate and low sp$^3$ amorphous carbon features are observed at ≈1500 cm$^{-1}$ [13]. Post annealing, three peaks are observed. The ≈1595 cm$^{-1}$ and ≈2268 cm$^{-1}$ peaks are strongly correlated with high sp$^3$ amorphous carbon composite[13], while the ≈1325 cm$^{-1}$ peak represents the partially recovered top membrane layer[14].

Figure S2 (b) shows the Raman peak of a 110 nm thick, ≈250 nm overgrowth diamond membrane grown in high temperature condition for 40 h. The sample is named as "N doped membrane" because of the in-situ doping of $^{15}$N during overgrowth. The Raman linewidth of this membrane (1.486(14) cm$^{-1}$) is slightly broader than the low temperature one. It is likely that this slight broadening of 0.11 cm$^{-1}$ originates from the higher heterogeneity of the crystal quality due to slight change in growth conditions from the delta doping and higher synthesis temperatures.

## 2.3 Membrane thickness and growth rate characterization

In this work, we applied two different approaches to ascertain the thicknesses of the membranes. For the growth rate analysis and thickness-insensitive applications, we used a profilometer (Bruker DektakXT) to measure the height difference between the diamond membranes and the carrier wafers supporting them. For thickness-sensitive applications, we use an ellipsometer (Horiba UVISEL2) to optically monitor the thickness during etching processes.

For the profilometry characterizations the height limit is set to 6.5 µm and the typical scanning length is set to 500 µm with scanning time ≈200 s. The scanning trajectory covers the full length of the membrane for a more accurate value. This measurement is launched after the HSQ annealing and before the "Ar/Cl$_2$" etching. According to the characterization results, transferred membranes without overgrowth are 340(5) nm thick. The thicknesses of the 40 h overgrown membranes are 680 nm to 740 nm (low growth temperature of 500 °C) and 570 nm to 600 nm (700 °C growth temperature), resulting in growth rates of 9.3(8) nmh$^{-1}$ and 6.2(4) nmh$^{-1}$, respectively (growth rate uncertainty from multiple membranes measured). For thickness-insensitive applications, the ICP etching recipe for each membrane is calculated from measured initial thickness, the target thickness, and the ICP etching rates (Table S2).

We also used profilometry characterization to ascertain the overall membrane height and flatness. A scan across a $^{15}$N-doped diamond membrane after the transfer is shown in Figure S5 (e). The result indicates an average membrane thickness of 603(10) nm before the back etching. This value shouldn't be affected by the back etching process as the ICP etching rate is uniform within the area of the membrane. Due to the angle between the membrane edge and the scanning direction, the length of the membrane in this figure seems to be longer than 200 µm. Since the scanning result includes some local height variation on the membrane, the real flatness should be less than 10 nm, which indicates a very uniform thickness across the whole membrane area.

This excellent uniformity is crucial for future device fabrication and potential applications.

For thickness-sensitive applications, we used the ellipsometry scanning to the diamond membranes. This technique was firstly applied in nanophotonic integration studies[15]. The scanning area is set to 85 µm × 35 µm and is carefully aligned to the membrane surface. The scanning step is set to ≈0.05 eV with long probing time at each step (≈1 s) to reduce the fitting error. To fit the diamond layer, we use Cauchy's equation for transparent material

$$n(\lambda) = A + \frac{B \cdot 10^4}{\lambda^2} + \frac{C \cdot 10^9}{\lambda^4}, \qquad (1)$$

where $A$ = 2.378, $B$ = 1.300 and $C$ = 0.000, extracted from[16]. By using this model, we are able to get a ≤4 nm height uncertainty for ≤50 nm membranes.

## 2.4 HSQ fluorescence analysis and vapor HF treatment

As the adhesion layer between diamond membranes and carrying wafers, the HSQ introduces extra fluorescence under laser excitation. These additional counts mostly span from 560 nm to 850 nm and can affect the PL measurements to some extent, depending on the types of color centers being measured.

The resonant PLE measurements are not affected due to their narrow optical linewidths, which allows for the use of exceptionally low excitation power (≤100 nW) and realize a large contrast of ≈30 against the background and APD dark counts. Therefore, for experiments and technologies that rely on resonant interactions, such as those proposed for quantum networking, the HSQ background will have minimal impact on device performance. The off-resonant photoluminescence measurements on group IV color centers (such as GeV$^-$) are mainly used for identifying the center locations, and are not critical for their performance in quantum networking. Due to the high Debye-Waller factors, group IV centers are typically measured with a narrow bandpass filter, which filters out most of the fluorescence from HSQ. With 10 mW excitation (532 nm) and appropriate filter sets, the HSQ fluorescence is ≈10 kHz to 20 kHz, which is less than the ZPL counts from individual centers (≈20 kHz to 40 kHz, shown in Figure 3(b) of the main text). Therefore, the presence of HSQ does not fundamentally hinder the optical characterization of GeV$^-$ centers or preclude the usage of GeV$^-$ in quantum networking applications, the primary technology driver for group IV color centers. Despite its limited impact, the HSQ background can be further reduced as wish via membrane suspension (≈2 kHz to 5 kHz) or vapor HF undercut (≈0.5 kHz to 1.5 kHz).

For NV$^-$ centers, the usage of a single longpass filter (561 nm) leads to a relatively high HSQ counts (300 kHz to 500 kHz under 10 mW green laser excitation), which hinders the PL and ODMR measurements. Therefore, NV$^-$ measurements benefit from reducing the HSQ background counts. In this work, we combined the suspended membrane fabrication (discussed in section 1.8) and the vapor HF treatment (discussed below) to reduce the total background to 1 5kHz to 30 kHz, which is comparable to the bulk diamond background. Other potential methods to reduce the HSQ background, such as HSQ patterning before transfer, usage of tri-acid clean to oxidize the HSQ layer, or material substitution with less optical fluorescence, will be explored in future studies.

The vapor HF treatment was used optionally to further minimize the background fluorescence from the HSQ underneath the suspended membrane region. During the vapor HF treatment, the chamber pressure is set to 24.5 Torr, with H$_2$O flow at 18.9 mLmin$^{-1}$ and HF flow at 40 sccm. Figure S5 (a) shows a bare carrier wafer after trench fabrication, and Figure S5 (b) shows one of the trenches after the HSQ deposition, annealing, and vapor HF. The image of the wafer prior to vapor HF can be seen in Figure S4 (d), where a colorful interference gradient can be observed. By carefully tuning the process duration (2 s), we are able to remove most of the HSQ on the sample, while leaving the thermal oxide layer almost untouched. Thus the HSQ has a negligible contribution to the background fluorescence under laser excitation. The sample shown in Figure 3 (a) of the main text received an excess HF treatment (10 s). The zoomed in image is shown in Figure S5 (c), where an undercut region (rainbow color) with ≈18 µm width can be seen, as well as some undesired surface patterns on the SiO$_2$ - Si carrier wafer (dark dots). Those unwanted effects are largely suppressed by shortening the process duration to 2 s, as seen in Figure S5 (d). The undercut width is ≈5 µm with no visible damage on the carrier wafer.

## 2.5   Dopant and isotope characterization via SIMS

The quality of the isotopic enrichment (replacement of naturally abundant $^{13}$C with $^{12}$C) is characterized via SIMS (EAG labs) and a typical carbon isotope scan is shown in Figure S9 (a).

We reproducibly achieve 99.99 at.% $^{12}$C diamond during overgrowth, a two orders of magnitude reduction compared to natural abundance as shown in Figure S9 (a). This is currently limited only by the isotopic enrichment of the precursor CH$_4$ gas.

Nitrogen doping effectiveness as a function of nitrogen precursor gas flow rate was also characterized via SIMS (EAG labs). Characterization of this was done via multiple scans on multiple delta doped samples with specific nitrogen doped layers (multiplicity for statistics). Fitting of the Gaussians (given the delta-doping layer thickness, a Gaussian serves as a valid approximation for the nitrogen distribution) was used to quantify the nitrogen density, which was only restricted by the depth distribution of these delta-doped layers. The FWHM is then used as an approximation of the delta-doping layer thickness, whereas the amplitude is used as an approximation of the atomic density (cm$^{-3}$) within that layer. A typical characterization run for a $^{15}$N triple delta-doped diamond is presented in Figure S9 (b). It remains to be noted that we also regularly characterize the presence of $^{14}$N to look for contamination, however, this value is regularly below the SIMS detection limit of about $2 \times 10^{-15}$ cm$^{-3}$.

# 3 group IV center characterization

## 3.1 Experimental setup for group IV center measurements

The membrane samples are mounted in a closed-loop cryostation (Montana S200) for low temperature measurements. The sample temperature during the measurements is 5.38 K as ascertained by a sensor (Lakeshore Cryotronics) installed directly on the sample mount. The position of the membrane is controlled by three closed-loop piezo positioners (Attocube ANC 350). The incident light beams are navigated by a fast steering mirror (Newport FSM-300) controlled by an analog output device (National Instruments, PCIe-6738). For the PL measurements, we use a 532 nm continuous wave (CW) laser (Lighthouse Photonics Sprout-G) as the excitation source. For photoluminescence excitation (PLE) measurements of GeV$^-$, the excitation laser is generated by a wave mixing module (AdvR Inc.) combining a tunable CW Ti:Sapphire laser (M Squared Solstis) and a monochromatic CW laser in telecommunication band (Thorlabs, SFL 1550P). We use this laser to scan across the GeV$^-$ ZPL wavelength (C line) and collect phonon sideband counts. At the collection ports, we use a 50 : 50 cube beam splitter (Thorlabs Inc. BS013) to divide signals to two branches for counts collection and spectroscopy measurements. A single photon counting module (SPCM) (Excelitas Technologies) is mounted to the counting beam, while a 1200 grmm$^{-1}$ spectrometer (Princeton Instruments, SpectraPro HRS) is fiber coupled to the spectra beam. The camera of the spectrometer (Princeton Instruments, PIXIS) is maintained at $\leq -70$ °C to minimize background counts. The spectrometer branch can also be connected to another SPCM to perform autocorrelation measurements. We use a combination of bandpass filters (Semrock FF01-615/24-25, Semrock FF01-600/14-25) for GeV$^-$ PL measurements. When performing PLE measurements, we collected the phonon sidebands via double filters (2× Semrock FF01-647/57-25). All lenses in the confocal setup are achromatic doublets with AB coatings (Thorlabs Inc.) to reach maximum transmission efficiency over broadbands. The PL map obtains a very low background fluorescence (≈1500 Hz) even under 10 mW excitation, as shown in Figure 3 (b) of the main text. The optical setups for SiV$^-$ and SnV$^-$ characterizations are almost identical, except for the bandpass filter (Semrock FF01-740/13-25 for SiV$^-$ and a Semrock FF01-615/24-25 for SnV$^-$). Since GeV$^-$ centers are also present in the PL map of SnV$^-$ centers, we use the spectrometer to identify the center type before proceeding with further characterizations.

## 3.2 GeV$^-$ creation efficiency

We estimate the creation yield of GeV$^-$ by counting the number of fluorescent centers in a certain area. Given the good signal-to-background ratio of ≈30, any resolvable centers when counting each 20 µm by 20 µm PL map were included in our counting. The average number of centers per areal spot is 53(3) (≈0.13 µm$^{-2}$), yielding a creation yield of 6.6(4)% from the implantation dose ($2 \times 10^8$ cm$^{-2}$≈2 µm$^{-2}$). This yield is higher than the previous reported value[17]. We note the high signal-to-background ratio can improve the visibility of darker GeV$^-$ centers, which potentially increase the estimated yield.

## 3.3 Spectra of Group IV centers and GeV auto-correlation measurements

The GeV$^-$ membrane sample were co-doped with Si$^+$ and Sn$^+$ with the same density and target depth. Figure S6 (a-c) shows the representative spectra of GeV$^-$, SiV$^-$ and SnV$^-$ centers under 5.4 K. Like GeV$^-$ centers, the ZPL peaks of SiV$^-$ and SnV$^-$ are comparable with the bulk diamond values[18,19]. The A and B peaks of SnV$^-$ are not resolvable due to large spin-orbit coupling and limited phonon population under such a low temperature. More in-depth studies of SiV$^-$ and SnV$^-$ centers are left for subsequent work to explore.

Off- and on-resonance measurements are performed on GeV$^-$ centers, as shown in Figure S6 (d). Both curves exhibit anti-bunching features with $g^{(2)}(0)$ well below 0.5, originated from their single center nature. During PL auto-correlation measurements, the off-resonance laser (532 nm) is operated at a slightly lower power (3 mW) to avoid bunching effects. The PLE autocorrelation measurement indicates a clear Rabi oscillation pattern induced by an on-resonance excitation.

The reasons for the non-zero $g^{(2)}(0)$ values (0.17(2) for PL and 0.19(5) for PLE) depend on the specific type of measurements. For PL autorrelation measurements, we observed a slowly decreasing ZPL signal with respect to the measurement time. This can be explained by a gradually changing surface termination with the presence of the excitation laser, which affects the preferred electronic configuration of GeV$^-$ centers. As a result, the color centers may experience more times in their optically dark state, slowly quenching the overall signal. In the future, a more rigorous surface termination scheme, such as a tri-acid clean or oxygen termination can be applied to improve the GeV⁻ stability. For the PLE autocorrelation measurements, the average ZPL linewidth is ≈2.4 times wider than the single scan values. With the frequency of the excitation source fixed, the spectral diffusion causes considerable fluctuations of phonon sideband counts, raising the PLE $g^{(2)}(0)$ value. Future optimizations on the linewidth of the GeV⁻ emissions can effectively enhance the average signal and suppress the $g^{(2)}(0)$.

### 3.4 GeV⁻ emission linewidth and spectral jump analysis

The optical linewidth of GeV⁻ centers approaches the lifetime limit by a factor of 3 (6) during single (average) scans. While the synthesis of optimized narrow linewidth color centers is beyond the scope of this work, multiple strategies can be introduced to overcome the extrinsic broadening factors. To begin with, the optical coherence is impacted by thermal effects via electron-phonon interactions, as modeled and characterized in[20]. Therefore, narrower linewidths may be achievable at lower characterization temperatures. Additionally, some inhomogeneous broadening is attributed to the implantation-induced damage[21]. In theory, this can be minimized by doping Ge in-situ during the overgrowth process[22], which is left for subsequent studies to explore. Furthermore, optimized surface terminations should contribute to reducing surface charge related noise. We excluded the optical power broadening in this work by characterizing the saturation power and tuning the excitation to be slightly lower than this saturation point before every linewidth measurement.

We also observed the switching phenomena for GeV⁻ in most membrane samples during PLE measurements. A long term scan (25 min) of a single GeV⁻ is shown in Figure S6 (e). This center is located in a 20 h membrane sample with only Si⁺ and Ge⁺ implantation. The origin of the switching is currently under investigation.

### 3.5 GeV⁻ zero phonon lines distribution

To evaluate the inhomogeneous broadening of the implanted GeV⁻ centers, we included an electronic grade diamond as a reference sample during the Ge⁺ implantation and subsequent annealing. Therefore, the reference diamond received identical ion implantation and annealing steps. Under the same cryostation temperature (5.4K), we measured a total of 75 GeV⁻ centers in diamond membranes and 30 centers in the reference sample, shown in Figure S6 (f) left and right. If all centers are included in deriving the inhomogeneous broadening, we find a standard deviation of $\sigma = 1.16$ nm for the membrane and $\sigma = 1.52$ nm for the reference sample. However, these numbers are mostly influenced by ≈10% of the centers with large wavelength shifts (such as 607 nm). These shifts are possibly induced by local strain distortion from the implantation process, and can be found in both membrane and bulk cases. Moreover, these $\sigma$ values are expected to have a large uncertainty due to limited sampling. Following a similar analysis[23], confining the ZPL wavelengths of GeV⁻ centers to within a range between 601.5 nm and 603.2 nm (which includes ≈90% of the centers) may give a better value for comparison. Compared to the inhomogeneous broadening in the reference ($\sigma = 0.18$ nm), the values we obtained are $\sigma = 0.18$ nm for the reference diamond and $\sigma = 0.19$ nm for the diamond membranes. These similar inhomogeneous broadening values indicate comparable crystal environments between the diamond membrane and bulk diamond.

# 4 NV⁻ characterization

## 4.1 Experimental setup for NV⁻ characterization

The room temperature NV⁻ coherence measurements were performed in a custom-built confocal microscope with a 532 nm, free-space laser (Oxxius 532S-150-COL-PP) and a 100X, 0.9NA air objective (Olympus MPLFLN100X). The laser spot position is controlled with a fast steering mirror (Newport FSM-300) and a 4f lens pair. The laser is gated with an acousto-optic modulator (AOM) (Gooch & Housego) and PL is measured with a PDM photon counter (MPD PD-050-CTD-FC). Microwave pulses are generated by a SRS SG-396, amplified by a MiniCircuits ZHL-15W-422-S+, and gated by both a Mini-Circuits ZASWA-2-50DR+ RF switch and external amplitude modulation of the signal generator. Pulses are applied to the NV⁻ centers via a 25μm Al wire draped over the sample. The experiment timing is handled by a Swabian Pulse Streamer.

A static magnetic field of 15 gauss is applied at approximately 35° to the [100] sample surface. Given the orientation of the sample, the field is at a 10° angle to the [111] crystal axis.

## 4.2 NV⁻ coherence measurements and $T_2$ comparison with delta-doped results from previous studies

We use standard free induction decay (FID) and spin echo (SE) sequences to measure $T_2^*$ and $T_2$, respectively. Each sequence, shown in Figure S8, is composed of two branches, which measure both the $|0\rangle$ and $|1\rangle$ projections. After applying a $\pi/2$-pulse on the NV⁻, we allow the spin to evolve for a time $\tau$ (and apply a $\pi_y$ pulse at $\tau/2$ for the SE) before applying either another $+\pi/2$-pulse or a $-\pi/2$-pulse to project into the readout basis. The readout windows at the beginning of each laser pulse are binned as the readout signals, $S_{0,-1}$, and the readout window at the end is binned as a reference, $R$, which allows us to correct for slow drift over each measurement cycle[24]. All measurements are processed as

$$\text{Coherence} = \frac{S_0 - S_{-1}}{R} \cdot \frac{R[0]}{S_0[0] + S_{-1}[0]}, \tag{2}$$

where the second term accounts for absolute count discrepancies between readout at the beginning and end of the laser pulse that arise from the finite rise time of the AOM. Each pulse sequence is a different length due to sweeping $\tau$, but run for the same amount of time overall, necessitating normalization by the reference to cancel this variation. Ultimately this means that shorter-$\tau$ data are average over more instances than longer-$\tau$ data. We fit $T_2^*$ measurements to the following[25]

$$B + \exp\left[-(t/T2^*)^2\right] \cdot \sum_{i=1}^{2I+1} a_i \cos\left(2\pi f_i(t+\phi_i)\right), \quad (3)$$

Where $B$ is a background offset, $a_i$, $f_i$, and $\varphi_i$ are the amplitude, frequency, and phase of each sinusoid corresponding to each hyperfine peak, and $I$ is the nitrogen nuclear spin (1/2 for $^{15}$N, 1 for $^{14}$N). $T_2$ measurements are fit to

$$B + A\exp[-(t/T_2)^n], \quad (4)$$

Where $B$ is an offset, $A$ is the amplitude, and $n$ is either a free parameter or fixed at $3^{25}$, when the former does not fit.

Previous delta-doping work demonstrated $T_2$ times for deep NV$^-$ centers ($\geq$52 nm) on the order of $\approx$700 µs[3], with measured values dropping drastically ($\leq$100 µs) as the delta-doped layer was brought close to the diamond surface ($\leq$35 nm). Indeed, it is expected that surfacerelated decoherence is a main limitation for the $T_2$ times presented in this work[12]. The impact of surfaces is further amplified in membranes, as the color centers are in close proximity to two surfaces rather than one.

## 4.3 NV$^-$ strain measurements

The NV$^-$ center is sensitive to both strain and electric field ($\mathbf{\Pi}$) through spin state coupling, as shown in the Hamiltonian

$$H = (D + \Pi_z)S_z^2 + \Pi_x\left(S_y^2 - S_x^2\right) + \Pi_y(S_xS_y + S_yS_x). \quad (5)$$

Parallel strain/electric field ($\Pi_z$) shifts the effective zero-field splitting, while the perpendicular components ($\Pi_{x,y}$) split the upper and lower spin transitions[26]. However, there is a discrepancy in the strength of the couplings for each physical origin. The electric field susceptibilities $d_{\epsilon:\parallel,\perp}$ =0.35 HzcmV$^{-1}$ and 17 HzcmV$^{-1}$ [27], while the strain susceptibilities $d_{\sigma:k,\perp}$ =13.3 GHz and 21.5 GHz[28]. We thus expect that if local electric fields dominate, the splitting between the upper and lower transitions should be much larger than the shifts in zero-field splitting, whereas if strain dominates these effects should be of similar magnitudes.

Through zero-field ODMR measurements, we found in all but one case that the splitting was greater than or the same magnitude as the shifts. In Figure S7 (a) we present the shifts in zero-field splitting versus the NV orientation, as determined at non-zero field, to extract an upper bound on the strain in the membrane. If we assume the larger shifts of $\approx$1 MHz arise entirely from strain – an over-estimate, given the above discussion – we arrive at a strain of $< 10^{-4}$.

Multiple factors can contribute to the built-in strain. For instance, the dry transfer step might introduce in-plane strain due to approach angle and speed. This can be reduced by reducing the approaching speed of the PDMS stamp with a smaller angle. Additionally, although the HSQ layer collapses above 465 °C and potentially releases the strain during the annealing, some residual strain might still remain. Furthermore, due to different thermal expansion ratios between the diamond membrane and carrier wafers, a thermal-induced strain appears when the sample is cooled down to room temperature or at 5.4 K of PL characterizations. The direction and the magnitude of the strain depend on the material of the carrier wafers. Finally, some on-chip structures (such as trenches) can alter the strain distribution across the membrane.

## 4.4 NV$^-$ $T_1$ comparison

We compare $T_1$ population decay times for two NV$^-$ centers with an order of magnitude separation in coherence times to confirm that a reduction in $T_1$ is not responsible for the reduction in $T_2$. Figure S7 (b) shows the differential $T_1$ measurement for NVX ($T_2 = 13(1)$ µs) and NVY ($T_2 = 146(53)$ µs). Over 5 ms there is no observable increased decay for NVX, indicating $T_1$ plays no role in limiting coherence.

## 4.5 [N] and [NV$^-$] estimation

Both the overgrowth background [$^{14}$N] and in-situ doped [$^{15}$N] were determined via calibrated quantitative secondary ion mass spectroscopy (SIMS) characterization (EAG Laboratories). All SIMS derived error bars are standard deviation from a duplicate experimental characterization. The SIMS detection limit of [$^{14}$N] placed an upper bound on the background nitrogen contamination $4.5(2) \times 10^{15}$ cm$^{-3}$ (≤26(1) ppb). For the 0.06 sccm $^{15}$N$_2$ dopant precursor flow rate used for the generation of in-situ doped $\delta$-doped membranes, we obtain a [$^{15}$N] of $5.1(10) \times 10^{15}$ cm$^{-3}$ (30.8(57) ppb) with an EAG quoted $^{15}$N detection limit of $1 \times 10^{15}$ cm$^{-3}$ (5.7 ppb).

The nitrogen implanted membrane sample was subjected to a 48 keV, $2 \times 10^8$ cm$^{-2}$ [N] dose. At this energy, the straggle (standard deviation of implantation as approximated with SRIM) is 14 nm. Assuming the bulk of the nitrogen resides in a ≤30 nm thick region, the expected [N] from the implantation to be $6.7 \times 10^{13}$ cm$^{-3}$ (≈0.4 ppb) within the implanted region. Comparing this to the observed [$^{14}$NV$^-$] of $1.8 \times 10^{12}$ cm$^{-3}$ (0.01 ppb) obtained from a typical count of a 10 µm by 10 µm region of the 120nm thick membrane. This suggests an upper bound on the N-to-NV conversion efficiency at ≤2.5%. Realistically, in-grown background [N] are also being converted to [NV$^-$], causing an overestimation of this conversion efficiency.

Isotopic tagging of dopant species in the $^{15}$N $\delta$-doped membrane was used to provide further insight into this. From a PL survey of a 12 µm by 12 µm area of the 110 nm thick *delta*-doped membrane (post-irradiation and annealing), we noted a relative ratio of 7 : 11 for [$^{15}$NV$^-$] : [$^{14}$NV$^-$]. Knowing that the delta doped area is ≈2 nm thick, with [$^{15}$N] = $5.1(10) \times 10^{15}$ cm$^{-3}$, we can work out the approximate density of the [$^{14}$N] in the rest of the membrane—assuming a uniform conversion efficiency for both the $\delta$-doped [$^{15}$N] and background in-grown natural abundance [N] (this approximation can break down easily, but remains useful here). Once geometric effective layer thicknesses are considered, i.e., 2 nm thick portion of [$^{15}$N] vs a 110 nm thick ≈ [$^{14}$N] membrane, we end up with a normalized observed [$^{15}$NV$^-$] : [$^{14}$NV$^-$]$_{normalized}$ of 7 : 0.2, resulting in a background [$^{14}$N] of $0.124(5) \times 10^{15}$ cm$^{-3}$ (0.70(3) ppb).

| Species | Energy [keV] | Dose [cm$^{-2}$] | Target depth [nm] | Straggle [nm] |
|---|---|---|---|---|
| $^4$He$^+$ | 150 | $5 \times 10^{16}$ | 413 | 39 |
| $^{14}$N$^+$ | 48 | $2 \times 10^8$ | 60 | 15 |
| $^{28}$Si$^+$ | 58 | $2 \times 10^8$ | 40 | 11 |
| $^{74}$Ge$^+$ | 98 | $2 \times 10^8$ | 40 | 10 |
| $^{120}$Sn$^+$ | 150 | $2 \times 10^8$ | 42 | 8 |

**Table S1**: Implantation parameters and SRIM simulation results for diamond membrane formation and color center generation.

| Recipe name | Ar/Cl$_2$ | O$_2$/Cl$_2$ | O$_2$ | Al$_2$O$_3$ etching |
|---|---|---|---|---|
| ICP power [W] | 400 | 700 | 700 | 400 |
| Bias power [W] | 250 | 100 | 100 | 50 |
| Pressure [mTorr] | 8 | 10 | 10 | 5 |
| Cl$_2$ flow [sccm] | 40 | 2 | 0 | 0 |
| Ar flow [sccm] | 25 | 0 | 0 | 10 |
| BCl$_3$ flow [sccm] | 0 | 0 | 0 | 30 |
| O$_2$ flow [sccm] | 0 | 30 | 30 | 0 |
| Etching rate [nmmin$^{-1}$] | ≈73 | ≈177 | ≈175 | ≈63 (Al$_2$O$_3$) |

**Table S2**: Cl-based ICP etching recipe for diamond. The last recipe, Al$_2$O$_3$ etching, is used for hard mask (ALD Al$_2$O$_3$) removal during the membrane patterning.

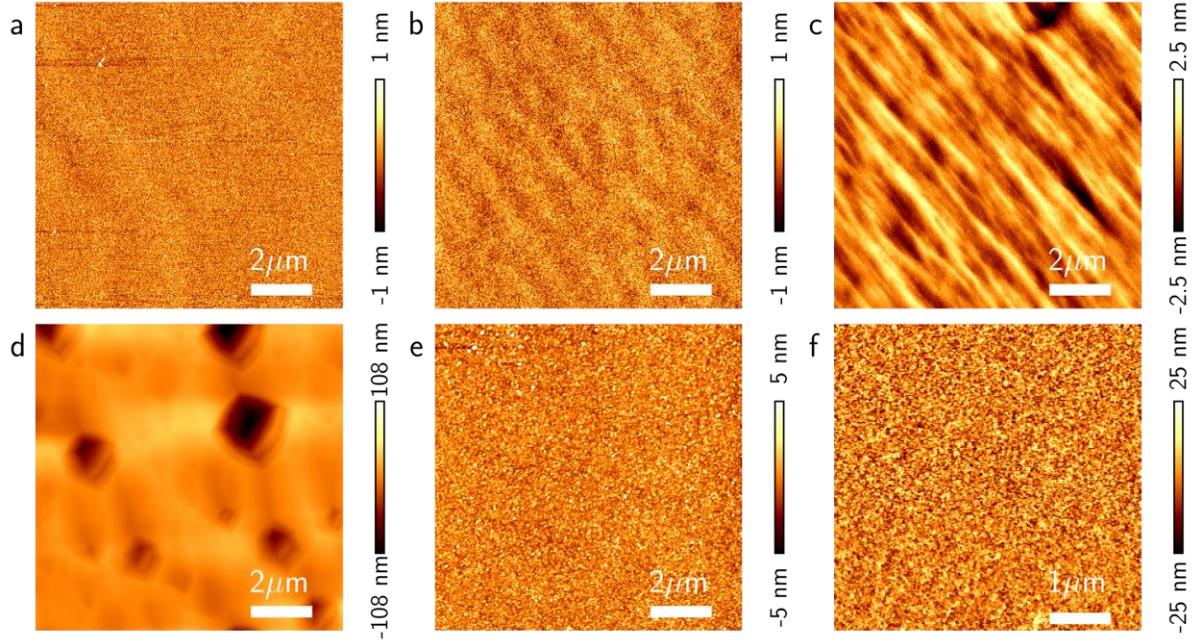

**Figure S1**: AFM characterizations during diamond membrane synthesis. (a) Fine polished diamond substrate shows $Rq$ = 0.30 nm. (b) The substrate after implantation and annealing with $Rq$ = 0.27 nm. (c) The 40h overgrown sample with 60min hydrogen etch. $Rq$ = 0.69 nm. (d) The 40 h overgrown sample with insufficient surface cleaning prior to the growth. $Rq$ = 18.6 nm. (e) Backside of the membrane after the dry transfer. $Rq$ = 1.17 nm. (f) Backside of the membrane after 30 s of oxygen ICP with no prior "Ar/Cl$_2$" etching. $Rq$ = 6.78 nm.

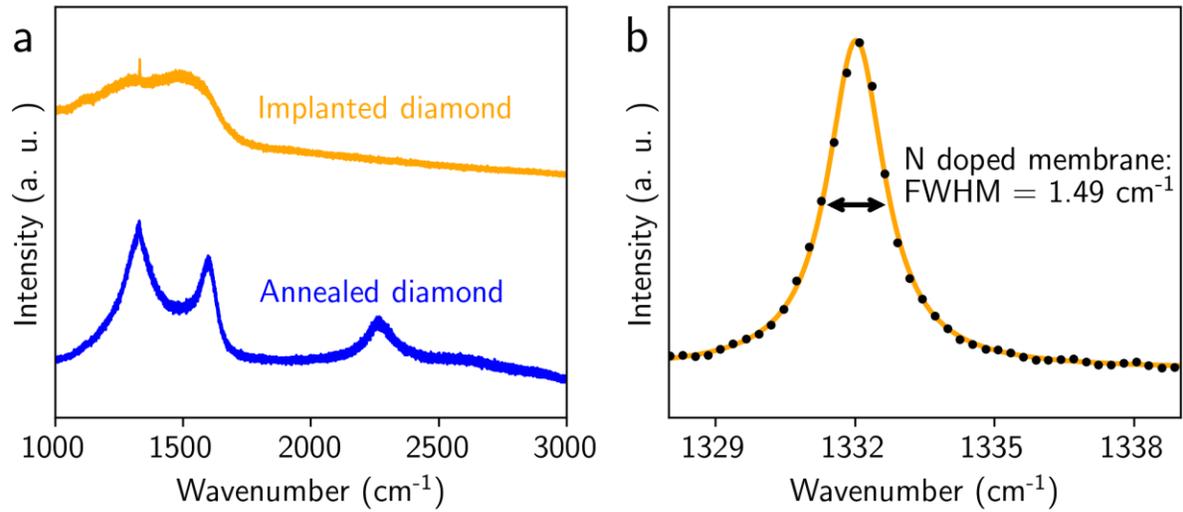

**Figure S2**: Additional Raman spectroscopy for membrane formation and overgrowth. (a) Orange curve: the implanted diamond substrate prior to annealing. Blue curve: the annealed diamond substrate. (b) The Raman peak for an isotopically purified, ≈250 nm overgrown membrane at 700 °C (40 h growth, back-etched down to 110 nm). The peak is at 1332.03(30) cm$^{-1}$, with linewidth 1.486(14) cm$^{-1}$.

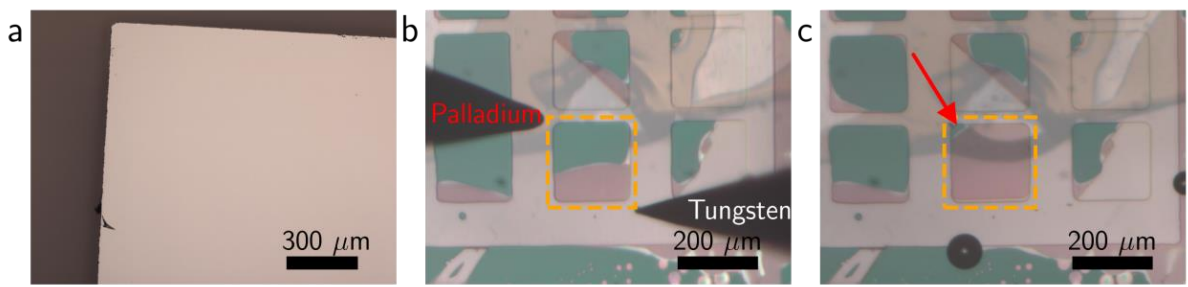

**Figure S3**: Microscope images for annealed and undercut diamond membranes. (a) The diamond membrane after He$^+$ implantation and subsequent annealing. The sample exhibits uniform light pink reflections, which can be interpreted as interference effect between a transparent top surface and a reflective graphitic underlayer. (b-c) A half- (almost-) undercut membrane (dashed orange square) at the EC etching step. Both images were taken when the membrane was in DI water. The left (right) electrode in (c) represents the palladium (tungsten) tip. The red arrow in (d) points to the membrane tether. The opaque graphitic layer exhibits interference colors (green), while the undercut region is almost transparent. The plate on the upper right indicates a leftover tether where the membrane has been picked up, while the membranes on the top and right were by-caught during the transfer. Such by-caught phenomena were caused by undesired EC etching due to low resistance of tap water and has been largely suppressed since the DI water is applied instead (See the membrane on the left in both images).

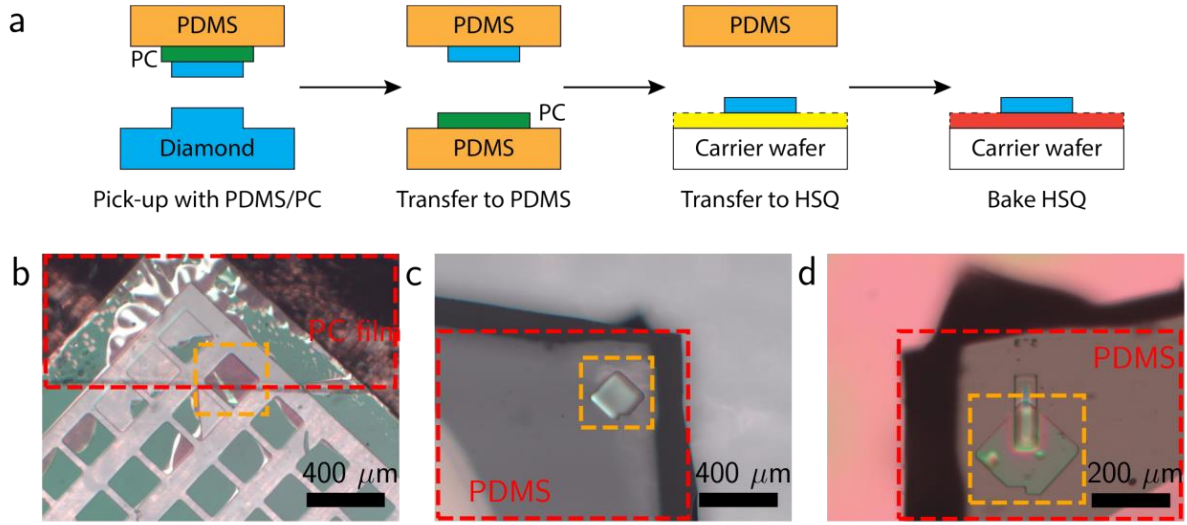

**Figure S4**: Deterministic diamond membrane transfer. (a) General transfer process flow. The diamond is flipped during the process in order to remove the original membrane afterwards. (b) The PDMS/PC stamp approaches the diamond membrane. The PC film (red dashed rectangle) is located under a PDMS stamp which spans beyond this image. The target membrane (orange dashed square) is linked to the diamond substrate via a small tether (See S3 (c)). (c) The membrane transferred to the PDMS-only stamp (red dashed rectangle) in the second step. (d) The membrane attached to a HSQ-coated trench on a thermal oxide wafer. The bumps under the membrane can be avoided by cleaner PDMS/PC stamp preparation in the future.

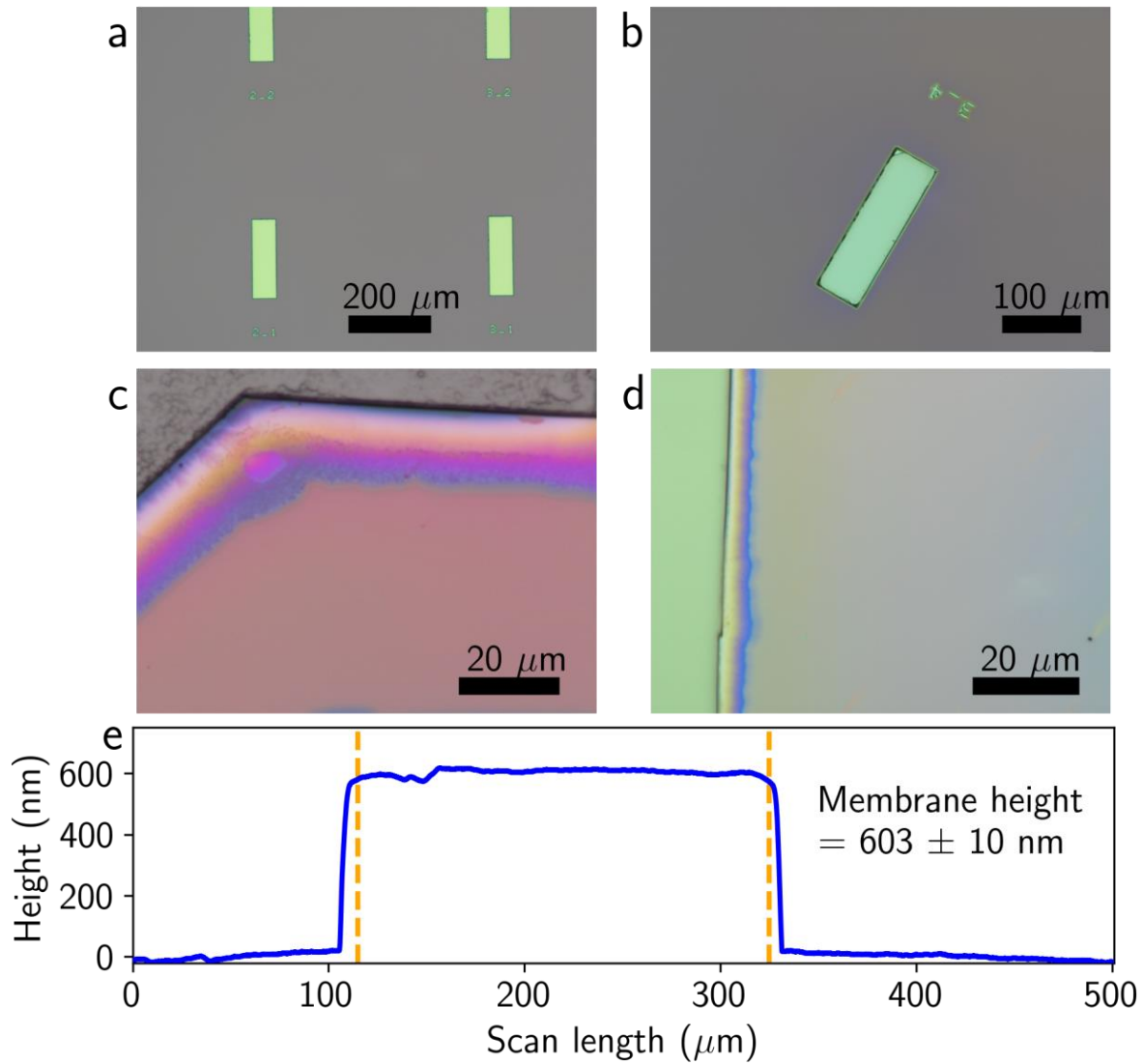

**Figure S5**: Carrier wafer preparation, vapor HF treatment and membrane thickness characterization. (a) 280 nm SiO$_2$–Si thermal oxide wafer with 5 µm deep trenches and markers. The silicon (green) exhibits different color from the oxide layer (gray). (b) A trench on the thermal oxide wafer after HSQ spinning, annealing and vapor HF treatment. The HSQ is mostly removed while the thermal oxide layer is barely untouched. (c-d) The undercut edge of the diamond membrane after 10 s (2 s) of vapor HF etching. (e) Profilometry characterization of the membrane thickness after transfer and before back etching. The region of the transferred membrane is indicated by the dashed orange lines. The average thickness for this $^{15}$N-doped membrane is 603 nm with a standard deviation of 10nm.

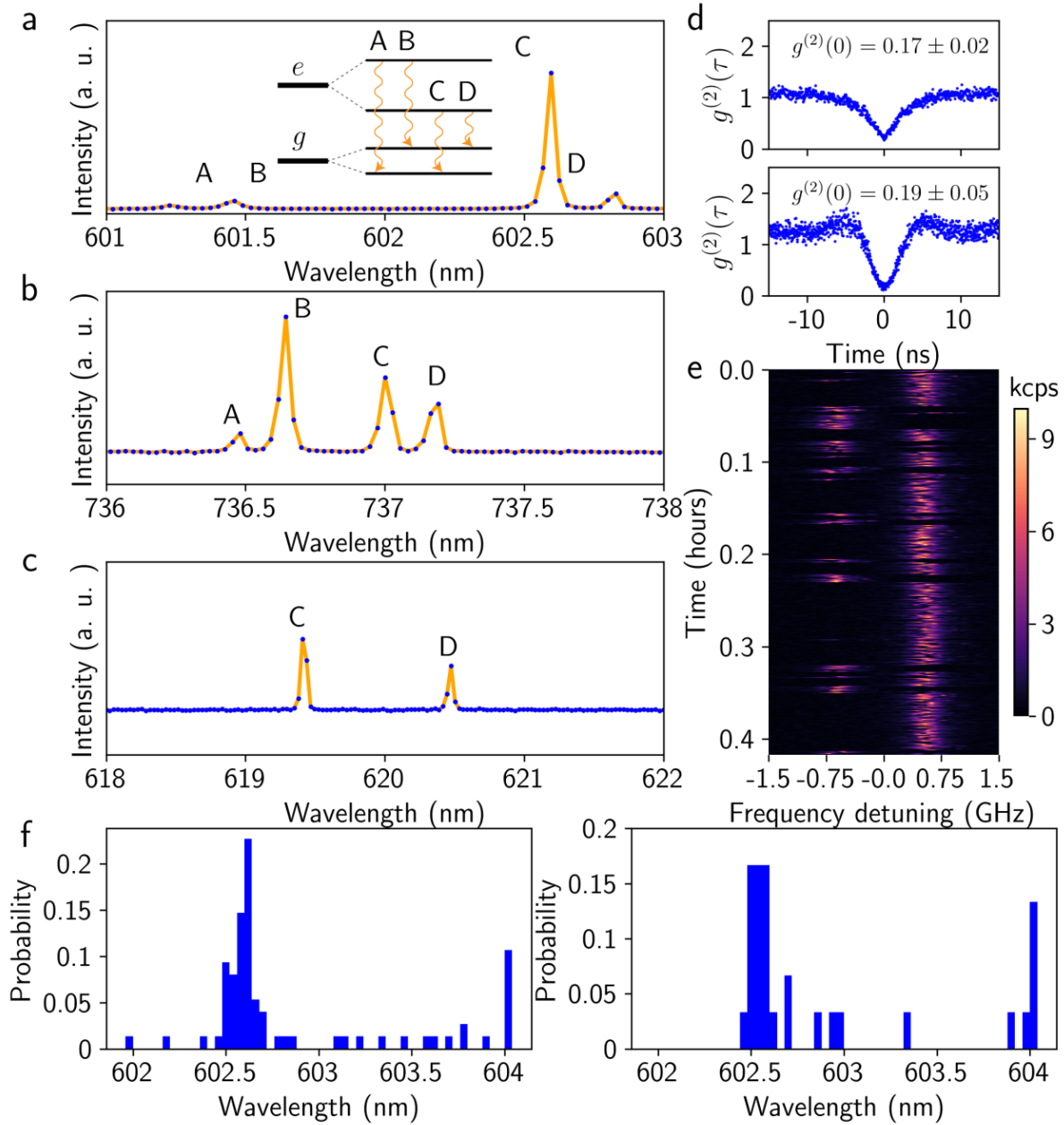

**Figure S6**: Additional optical measurements of group IV centers. (a-c) The spectra of a representative GeV (SiV⁻, SnV⁻) center measured under 5.4 K, with 10 mW 532 nm laser excitation. The small plot inside (a) represents the general energy levels of negatively charged group IV centers. (d) Off- (on)-resonance autocorrelation measurements of a GeV⁻ center. The excitation power of the off-resonance laser is 3mW, while the on-resonance laser is 40 nW. (e) A 25 min scanning of a spectrally switching GeV⁻ center. (f) ZPL distributions of GeV⁻ centers in the diamond membrane (left) and bulk diamond (right).

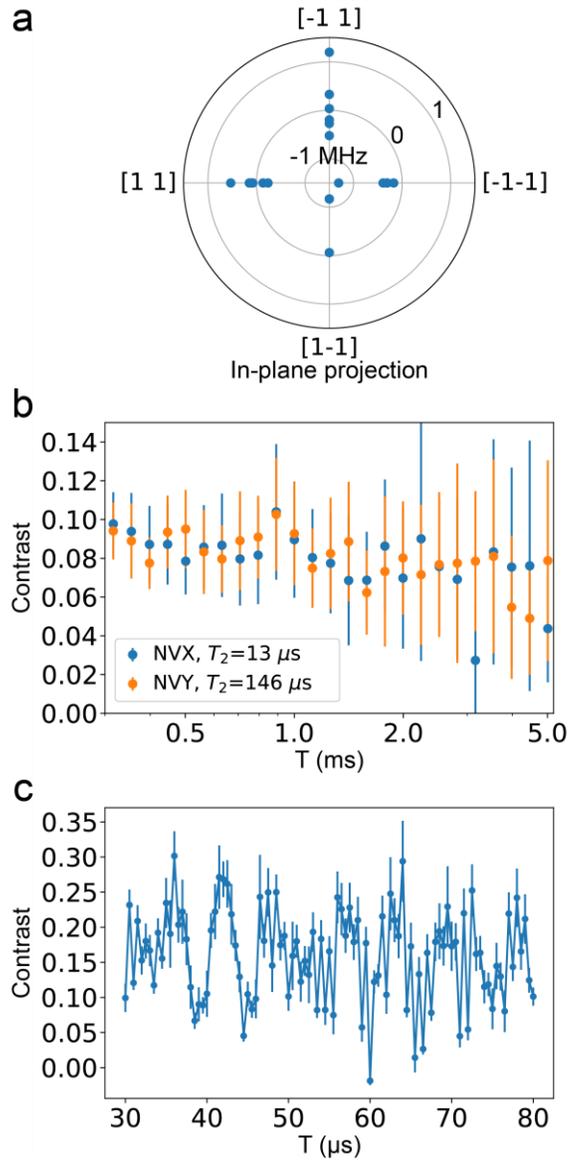

**Figure S7**: (a) Zero-field splitting of measured NV⁻ centers. (b) $T_1$ measurement of two NV⁻s with order of magnitude spread in $T_2$ times. (c) Finer resolution spin echo measurement on the NV⁻ presented in Fig. 4 (a), showing electron spin echo envelope modulation, which is aliased in the main text data.

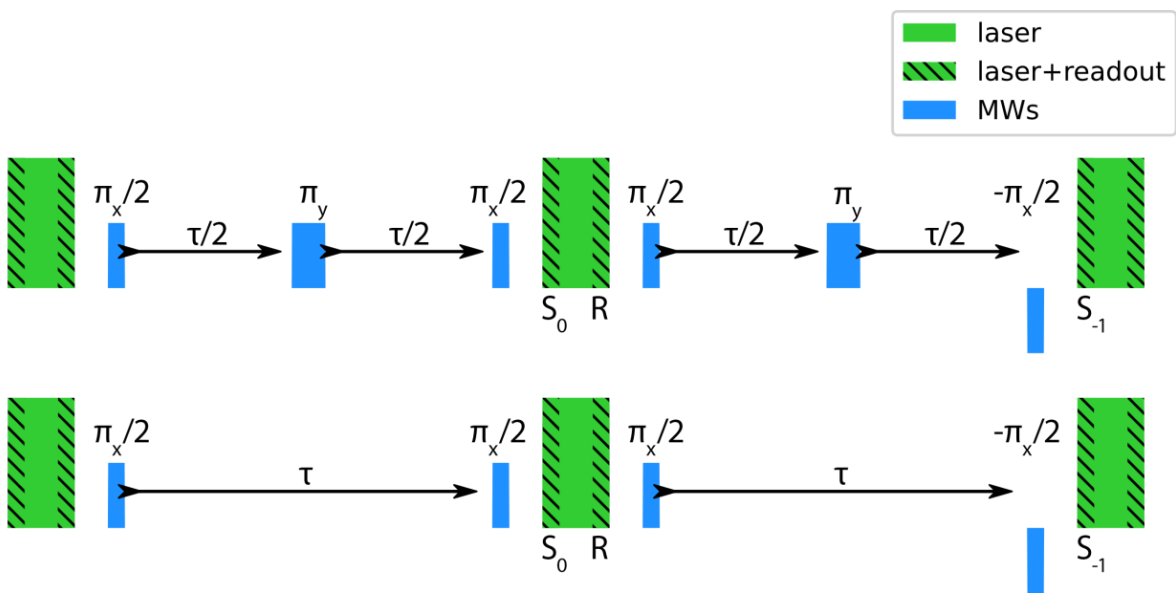

**Figure S8**: Coherence measurement sequences. Both the spin echo (a) and free induction decay (b) sequences have readout, initialization, state preparation, and state projection pulses. The spin echo sequence has a central $\pi$-pulse to refocus the spin.

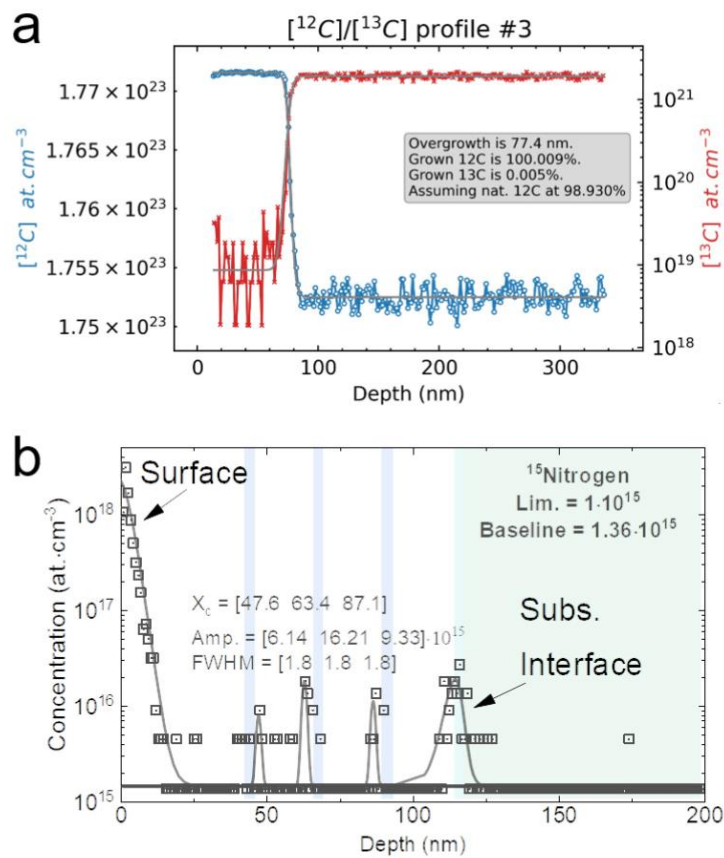

**Figure S9**: SIMS characterizations. (a) SIMS characterization of a typical isotopically enriched diamond growth using our standard overgrowth showing 99.99at.% $^{12}$C diamond. (b) Nitrogen doping effectiveness $^{15}$N (background contamination $^{14}$N not shown but done parallel to this) for a triple delta-doped overgrowth sample. Fitting of the Gaussians was used to quantify the nitrogen density. The FWHM is used as an approximation of the delta-doping layer thickness, whereas the amplitude is used as an approximation of the atomic density (cm$^{-3}$) within that layer. The detection limit is shown in graph. We also note a significant amount of nitrogen at the substrate/overgrowth and surface interfaces with ratios consistent of natural abundance of $^{15}$N in atmospheric nitrogen (measured via peak area ratio).